\documentclass[a4paper,fleqn,usenatbib]{mnras}

\usepackage{newtxtext,newtxmath}

\usepackage[T1]{fontenc}
\usepackage{ae,aecompl}

\usepackage{natbib}
\usepackage{graphicx,epstopdf}
\usepackage{amssymb}
\usepackage{amsmath}
\usepackage{amsfonts}
\usepackage{url}
\usepackage{mathptmx}
\usepackage{alltt}
\usepackage{listings}
\usepackage{lineno}
\RequirePackage{array}
\usepackage{makeidx}
\usepackage{multirow}
\usepackage{xspace}
\usepackage{longtable}
\usepackage[utf8]{inputenc}
\usepackage{float}
\usepackage{nameref}
\usepackage{color}
\usepackage{hyperref}

\title[Polarimetry of (3200)~Phaethon]{The phase-polarization curve of asteroid (3200)~Phaethon}

\author[Devog\`{e}le et al.]{
M. Devog\`{e}le$^{1}$\thanks{Partly based on data collected with 2-m RCC telescope at Rozhen National Astronomical Observatory, Bulgaria}
A. Cellino$^{2}$,
G. Borisov$^{3,4}$,  
Ph. Bendjoya$^{5}$, 
J.-P. Rivet$^{5}$, 
L. Abe$^{5}$, 
\newauthor
S. Bagnulo$^{3}$,
A. Christou$^{3}$, 
D. Vernet$^{6}$,
Z. Donchev$^{4}$, 
I. Belskaya$^{7}$,
T. Bonev$^{4}$,
\newauthor
and Yu.~N. Krugly$^{7}$\\
$^{1}$ Lowell Observatory, 1400 West Mars Hill Road, Flagstaff, AZ 86001 (U.S.A.) \\
$^{2}$INAF - Osservatorio Astrofisico di Torino, Pino Torinese, Italy\\
$^{3}$Armagh Observatory and Planetarium, College Hill, Armagh BT61 9DG, UK\\
$^{4}$Institute of Astronomy and National Astronomical Observatory, Bulgarian Academy of Sciences, 72, Tsarigradsko Chaussee Blvd., BG-1784 Sofia, Bulgaria \\
$^{5}$Universit\'e C{\^o}te d'Azur, Observatoire de la C{\^o}te d'Azur, CNRS, Laboratoire Lagrange UMR7293, Nice, France\\
$^{6}$Universit\'e C{\^o}te d'Azur, Obs. de la C\^ote d'Azur, UMR7293 CNRS Laboratoire Optique, Bv de l'Observatoire, CS 34229, 06304 Nice, France\\
$^{7}$Institute of Astronomy of Kharkiv National University, Sumska Str. 35, Kharkiv 61022, Ukraine\\
}

\date{Accepted XXX. Received YYY; in original form ZZZ}

\pubyear{2018}

\begin{document}
\label{firstpage}
\pagerange{\pageref{firstpage}--\pageref{lastpage}}
\maketitle

\begin{abstract}
A multi-colour phase-polarization curve of asteroid (3200)~Phaethon has been obtained during the December 2017 apparition by merging measurements taken at the observing station of Calern (France) and at the Rhozen observatory (Bulgaria).  All the observations were obtained in the positive polarization branch, the phase angle ranging from 36$^\circ$ to 116$^\circ$. The measured values of linear polarization are among the highest ever observed for a Solar system body. The covered interval of phase angle was not sufficiently extended to derive a firm determination of the $P_{\rm max}$ parameter, but this appears to occur at a phase angle around 130$^\circ$ and reaches more than 45\% of linear polarization. 
Phaethon is the parent body of the Geminid meteor shower, and the real physical nature of this object (asteroid or comet) has been a long-debated subject. Our polarimetric measurements seem to support the asteroid hypothesis with a phase-polarization curve similar to the asteroid (2)~Pallas, but further observations at smaller phase angles are needed to draw definitive conclusions. 
\end{abstract}




\section{Introduction}

The near-Earth asteroid (NEA) (3200)~Phaethon is a very interesting object in
many respects. It has an estimated size of about 5 km \citep{Hanusetal16}. Its most outstanding properties are:
\begin{itemize}
\item A very large orbital eccentricity, that produces a very 
small perihelion distance.
\item The fact of being a member of an unusual taxonomic class of low-albedo objects.
\item The fact of being the most likely parent body of the Geminid meteor shower.
\end{itemize}

The osculating orbital elements of Phaethon, taken from the NeoDys
web site\footnote{http://newton.dm.unipi.it/neodys2/} are listed in Table \ref{Table:oscel}.
They indicate that this object dynamically belongs to the Apollo
orbital class of the NEA population. During their orbital motion, these objects sweep regions 
inside the orbit of the Earth, but what makes Phaethon a quite peculiar Apollo is
its perihelion distance $q$ = 0.1399 AU, one of the smallest values ever 
found for an asteroid, and a consequence of its very
high orbital eccentricity of about 0.89.
\begin{table}
	\centering
	\caption{Osculating elements of (3200)~Phaethon at epoch 58200.0 MJD}
	\label{Table:oscel}
	\begin{tabular}{lccc}
		\hline
		& Value & 1-$\sigma$ variation & Units\\
		\hline
		$a$      & 1.27135  & 3.412e-10  & AU\\
		$e$      & 0.889933 & 6.935e-09  &    \\
		$i$      & 22.256   & 2.491e-06  & deg\\
		$\Omega$ & 265.226  & 2.038e-06  & deg\\
		$\omega$ & 322.173  & 2.108e-06  & deg\\
		$M$      & 38.954   & 8.531e-07  & deg\\
		\hline
	\end{tabular}
\end{table}

In terms of taxonomy, Phaethon was originally classified by \citet{Tholen85} as a member of the F class. This class, first introduced by \citet{GradieTedesco82}, included asteroids characterized by an overall lack of absorption features, and exhibiting a flat (F-class) reflectance spectrum down to the very blue region, where other classes tend to display some significant absorption bands. Based on IRAS thermal IR data, \citet{Tedescoetal2002} found the albedo of Phaethon to be 0.11 which is higher than the typical value for the F-class which ranges between 0.03-0.07. In the case of Phaethon, recent estimates based on joint modeling of IRAS and Spitzer data gives 0.12+/-0.01 \citet{Hanusetal16}. In the \citet{Tholen84} taxonomy, only 27 asteroids (3\% of all classified objects) were found to belong to this class. F-class asteroids are therefore uncommon. Based on their properties, \citet{Gaffeyetal89} suggested that they are most probably primitive objects, having reflectance spectra similar to those of CI1 - CM2 meteorite assemblages, and might likely include some organic compounds on their surfaces. 

Recently, however, the F taxonomic class has been removed from taxonomic classifications based on spectroscopic observations using modern CCD devices \citep{BusBinzel, Demeo}, because the blue part of the reflectance spectrum is no longer adequately covered. Objects previously classified as F have now been merged together with old B-class asteroids, which were previously distinct, in a unique class named B, which includes objects having featureless reflectance spectra, flat or slightly blueish in the wavelength interval between about 0.5 and 1 $\rm \mu$m. 

Since the epoch of its discovery in 1983, based on data collected by the IRAS satellite, (3200)~Phaethon has been considered to be the long sought parent body of the Geminid meteor shower, since its orbit was found to be practically coincident with that of 20 Geminid fireballs 
photographed in the 50s by the Prairie Network cameras. (see, eg \citet{Gustafson89} and references therein). The origin of several Geminid fireballs from Phaethon were also found by this author to be consistent with episodes of cometary activity of their assumed parent body. 

Based on all the above mentioned properties, a natural question arises: is Phaethon an asteroid or a comet? In this paper, we use the results of our recent polarimetric measurements of this object to provide some additional arguments to considered in this discussion. Though we can say that our new data are not yet sufficient to draw definitive conclusions, and waiting for some new data that we plan to obtain later this year during the next apparition of Phaethon, we already see evidence supporting one of the two alternative hypotheses concerning the real nature of this object.

\section{(3200)~Phaethon: asteroid or comet?}
\label{whatisit}

The following arguments can be used to support or contradict the two hypotheses about the real nature of Phaethon.

First of all, it is commonly accepted that the NEA population includes also objects which were originally born as comets, and achieved their current orbits after a complex dynamical evolution. A recent estimate of the relative abundance of cometary bodies among NEAs is $8\% \pm 5\%$ \citep{DeMeoBinzel08}.

Moreover, in recent years, the discovery of so-called main-belt comets, objects with typically asteroidal orbits, but found to exhibit transient cometary activity, has clearly shown that the classical distinction between asteroids and comets is probably not as sharp as we were led to believe in the past \citep{HsiehJewitt06}. In this respect, \citet{JewittHsieh06} presented observations of (155140)~2005~UD, an another small Apollo asteroid. Based on its orbital and spectroscopic similarities with Phaethon, taking into account also the different sizes of the two objects (2005 UD being only about 1 km in size) reached the conclusion that both Phaethon and 2005 UD might be separate pieces of a unique main-belt comet parent body that suffered disruption after reaching a Sun-approaching orbit.

It is important to note that, among objects originally classified as members of the old F-class, there is at least one major example of a body, (4015) Wilson–Harrington (see \citet{Bowelletal92}) that was originally considered to be an F-class asteroid, but was later found to exhibit cometary activity. No further activity has been detected again in recent years for Wilson-Harrington, suggesting that this object is comet very close to becoming extinct \citep{Fernandezetal2005}. Based on the observed CN spectral band emission at 0.388 $\rm \mu$m, which is typical of cometary activity, \citet{Chamberlinetal96} suggested that (3200)~Phaethon could be another inactive cometary remnant.

The possibility that asteroids originally classified as members of the old F-class display properties 
diagnostic of a cometary origin is exciting. A major point is that the old F-class asteroids can still be distinguished today based on their uncommon polarimetric properties. In particular, the phase-polarization curves of these bodies are characterized by unusually low values, of about 17$^\circ$ according to \citet{Belskayaetal17}, of the so-called polarimetric inversion angle, that is the value of phase angle at the point of transition from the negative polarization branch to the positive polarization branch. In other words, the inversion angle marks the transition from the interval of phase angles for which the measured plane of linear polarization is parallel to the Sun-target-observer plane (the scattering plane) to the interval of phase angles for which the plane of linear polarization becomes normal to the scattering plane \citep{Belskayaetal05, Cellinoetal15, Belskayaetal17}. For ``regular'' asteroids, the value of the inversion angle is larger than 18$^\circ$ and in the vast majority of cases it is around $20^\circ$ or larger \citep{Belskayaetal05, MNRAS2}. 
Interestingly, polarimetric observations of two comets, 2P/Encke \citep{Boehnhardt08} and  133P/Elst-Pizarro \citep{Bagnuloetal10} showed that their nuclei exhibit a polarimetric behavior characterized by low inversion angles, similar to those of known F-class asteroids. This supports previous suggestions \citet{KolokoloJockers} 
that a low inversion angle of polarization could be diagnostic of a cometary surface.

These are the main reasons why we decided to perform a campaign of polarimetric observations during the last apparition of Phaethon.The simple idea was to obtain a phase-polarization curve, to be compared with those of typical F-class asteroids, to check whether they share the same properties, in support a likely cometary origin of Phaethon. These observations, described in Section \ref{sec:obs}, are discussed in Section \ref{sec:discussion}. This apparition was also a unique opportunity to obtain measurement at a wide range of phase angle which allow to extend the phase coverage for this type of object and to compare them with either main belt asteroid measurements at lower phase angle and other NEOs.

On the other hand, there are arguments suggesting that (3200)~Phaethon could be more likely an asteroid, rather than a comet. First of all, the very belonging of Phaethon to the F class is not completely certain. Although \citet{Tholen85} assigned Phaethon to the F class based on a clear absence of an UV absorption feature, \citet{CochranBarker84} had previously reported a reflectance spectrum of Phaethon exhibiting a definite UV absorption feature. 

Later, \citet{Halliday88} pointed out that observations of Geminid fireballs suggested bulk densities between 0.7 and 1.3 g/cm$^3$, lower than the typical densities of meteorites, but higher than the values generally inferred for fireballs. This would suggest that Phaethon is not a (typical) comet. In this respect, it is important to note also that \citet{DeLeonetal10}, based on a comparison of the reflectance spectra, suggested that (3200)~Phaethon could be a fugitive member of a dynamical family whose parent body is the large main belt asteroid (2) Pallas. These authors noted that dynamical evolutionary paths possibly linking Phaethon with Pallas do exist. However, the spectrum of Phaethon is generally bluer than that of Pallas. The spectral similarity seems to be stronger between Phaethon and some small members of the Pallas family,  something that the above authors generically explained as a consequence of size-dependent surface effects. Finally, these authors also found that no other asteroid belonging to the modern B taxonomic class exhibits a comparable spectral similarity.

Finally, \citet{Hanusetal16} derived, using a sophisticated thermophysical model, a new value of 0.12 $\pm$ 0.01 for the geometric albedo of Phaethon. This value is much higher than previous estimates based on thermal IR IRAS data, and may be too high for a typical comet (just for a comparison, the geometric albedo derived for comet 67P/Churyumov-Gerasimenko from Rosetta observations is 0.062 $\pm$ 0.02, according to \citet{Ciarnielloetal15}). 
A value of 0.12 for the geometric albedo is rather high also for low-albedo asteroids belonging
to both the old F-class and the modern B-class. We note, however, that, in this respect, (2) Pallas has also a relatively high albedo value 
of 0.145 according to \citet{ShevTed}. Such value, which is anomalous when looking at the relations between 
geometric albedo and polarimetric properties \citep{MNRAS1} is confirmed, however, by independent
thermal radiometry estimates \citep{Masiero11}.

In this situation, we carried out our campaign of polarimetric observations of Phaethon.

\section{The observations}
\label{sec:obs}

The observations were done at the Calern observing station (MPC 010) of the Observatoire de la C\^ote d'Azur (France) and at the Bulgarian National Astronomical Observatory (BNAO), Rozhen (MPC 071), Bulgaria. At Calern, the Omicron (west) telescope of the C2PU facility was used with the Torino Polarimeter (ToPol) mounted on its Cassegrain focus (F/12.5).  ToPol is a Wedged-Double Wollaston polarimeter \citep{Oliva}. Its optical configuration makes it possible to obtain 
simultaneously four images corresponding to rotations of the linear polarization plane of $0^\circ$, $45^\circ$, $90^\circ$ and $135^\circ$, from which the $Q$ and $U$ Stokes parameters can be measured in one single exposure. ToPol is described in \citet{SPIE2012}. A description of the instrument and the reduction techniques used for this instrument have also been described by
\citet{Dev_2017}.

Polarimetric observations at BNAO-Rozhen were performed using the 2-Channel-Focal-Reducer Rozhen (FoReRo2) \citep{Jockersetal2000} 
attached at the Cassegrain focus of the 2m Ritchey-Chr\'etien-Coud\'e (RCC) telescope. In polarimetric mode, FoReRo2 is equipped with a 
Wollaston prism, placed in the parallel beam of the instrument. Recently, the instrument was equipped with a retarder half-wave-plate. 
This makes it possible to easily rotate the light electric vector before the Wollaston prism by steps of 45$^\circ$, 
(by simply rotating the retarder by steps of 22.5$^\circ$), and to use the so called “beam swapping technique”, described 
in \citet{Bagnuloetal09}, to measure accurately the linear polarization of the observed source.

At Calern, the polarimetric observations were performed in four colours, using standard BVRI filters, as long as the target was 
sufficiently bright and high above the horizon to allow for multi-colour measurements. We remind that 
Phaethon was moving at a very fast rate, with its apparent brightness and elevation decreasing from night to night, as the
phase angle was steadily increasing. This means that during the last nights when it was still observable, we could only
perform observations in V and R colours, which require shorter exposure times. At Rozhen, all measurements were done in R light, only. 

\begin{figure*}
\centerline{
\includegraphics[width=75mm]{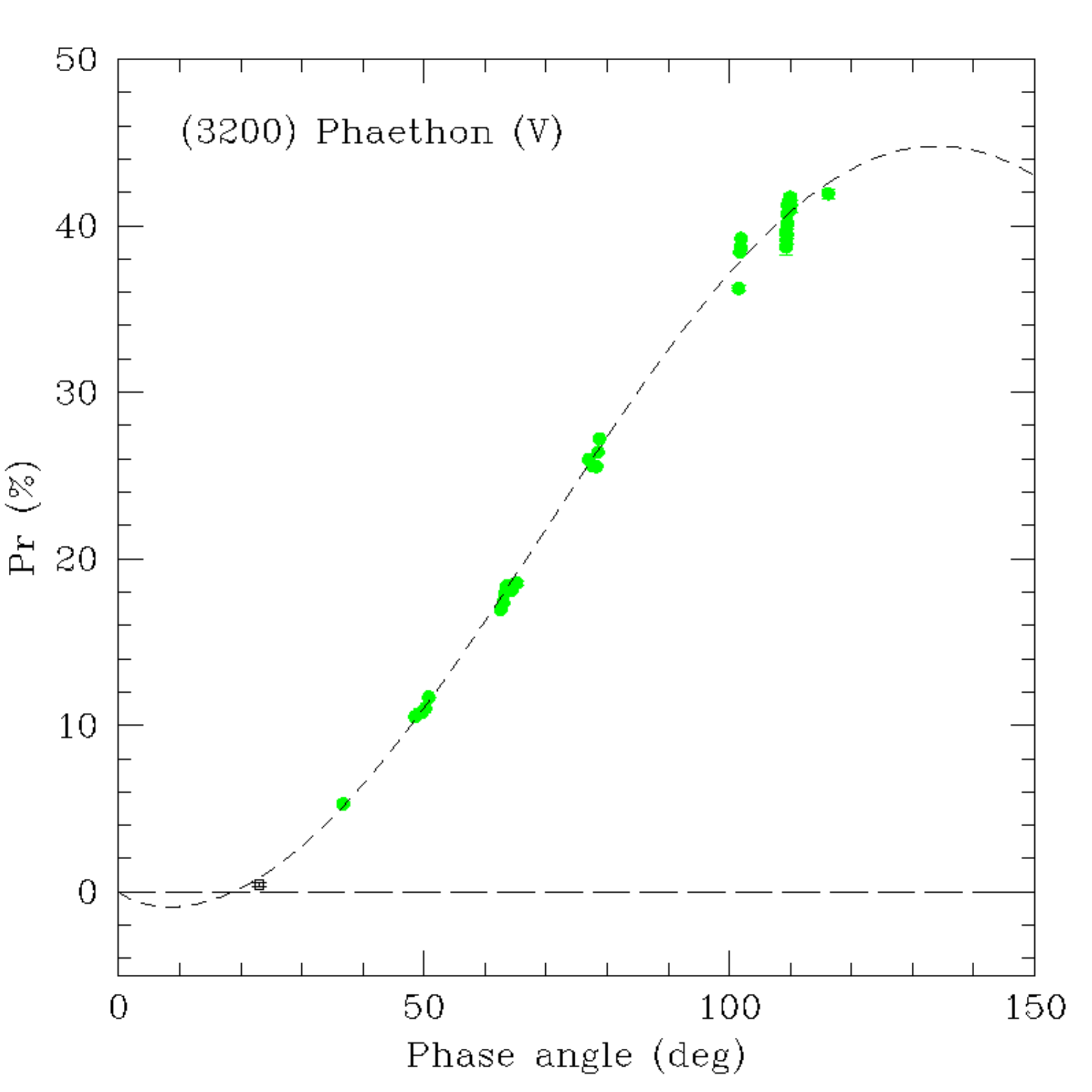}\hspace{1cm}
\includegraphics[width=75mm]{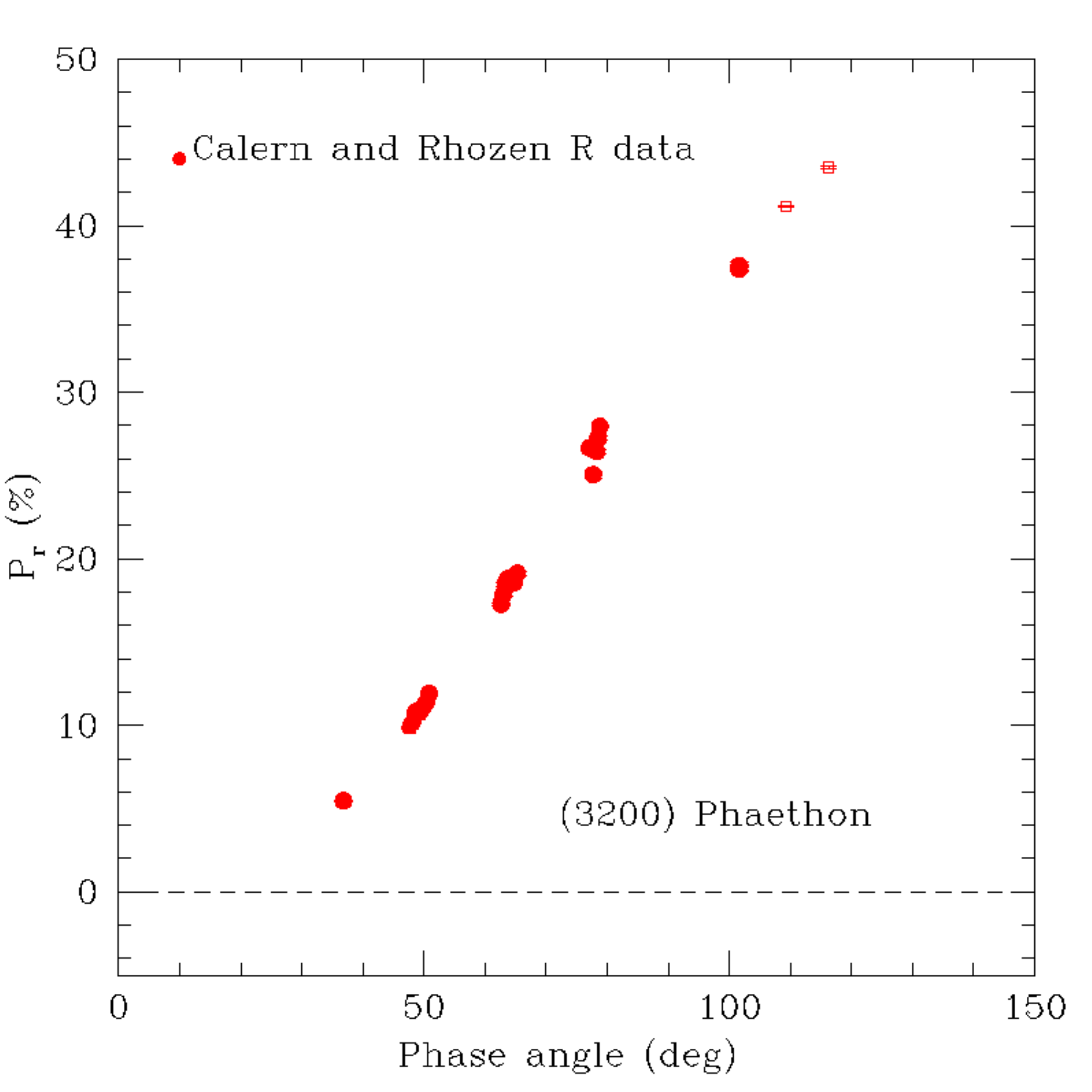} }
\caption{Left panel: Phase-polarization curve of (3200)~Phaethon in V light. New Calern data (green symbols) plus one data point 
(black square) previously
available in the literature. The best-fit curve computed using the trigonometric representation (see text) is also shown. 
Right Panel: Phase-polarization of (3200)~Phaethon in R light. Calern and Rozhen data are merged together. Rozhen measurements
are indicated by empty symbols, which are visible only at the highest values of phase angle, where they do not overlap
with simultaneous measurements done at Calern.}
\label{fig:phasepolcurveV+R}
\end{figure*}
\begin{figure}
	\includegraphics[width=\columnwidth]{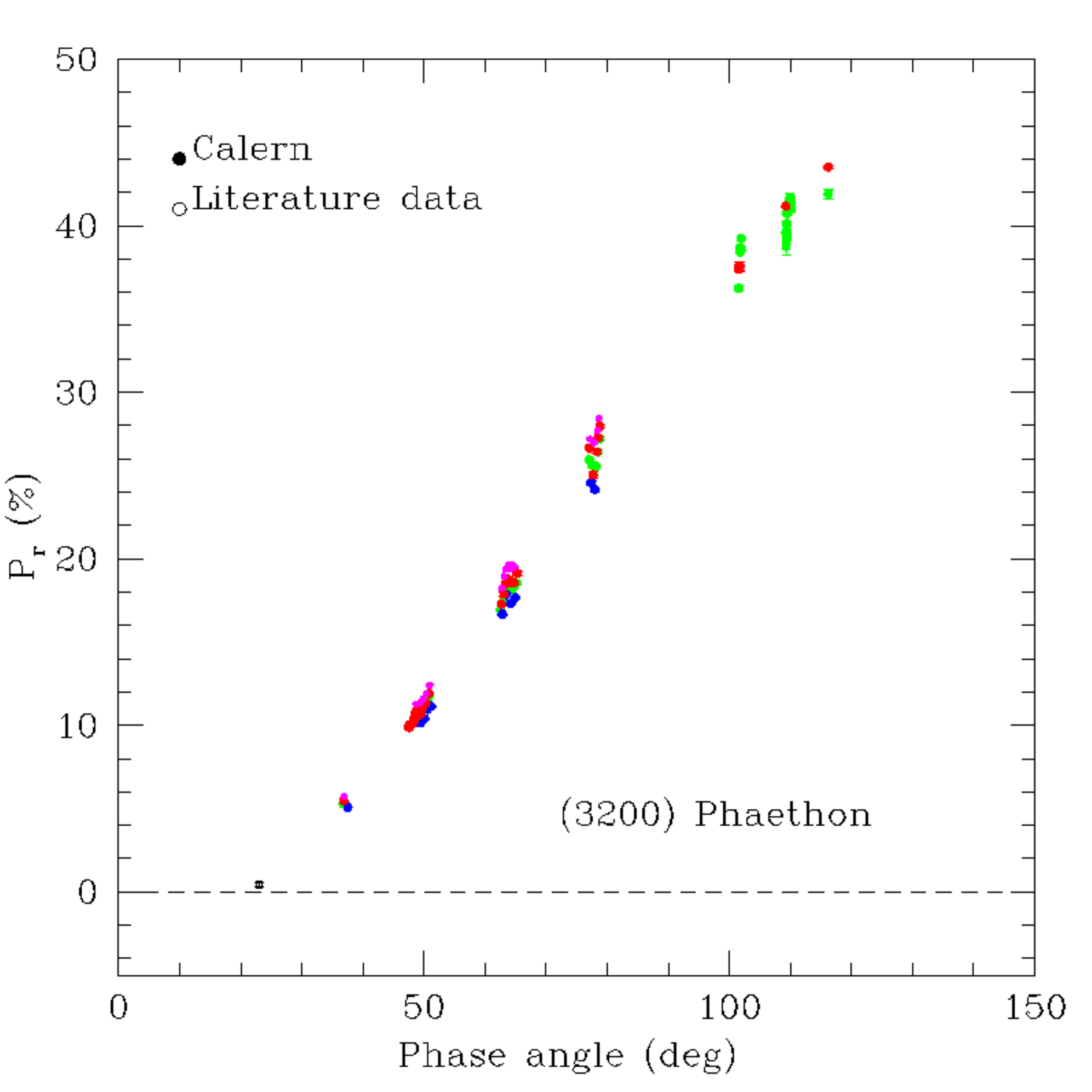}
    \caption{Calern and Rozhen polarimetric data for (3200)~Phaethon in BVRI filters. The corresponding symbol colours are blue, green, red and magenta, respectively. The single data point taken from the literature in V-colour is shown as a black point. Points corresponding to different colours tend to overlap in this plot, but there is a regular trend of increasing positive polarization going from B to I, as can be seen looking at the data shown in Table \ref{Tab:data}. Only V and R data were obtained at the highest phase angles, when the object was becoming fainter and located at small elevation above the horizon.}
    \label{fig:BVRICalern}
\end{figure}

\section{Discussion}
\label{sec:discussion}

Figures \ref{fig:phasepolcurveV+R} and \ref{fig:BVRICalern} show the results of our polarimetric campaign. The corresponding data are listed in Table \ref{Tab:data}. In the above Figures the nominal error bars of the measurements are shown, and they are often non-negligible, as shown in Table \ref{Tab:data}, but due to the large interval covered by the polarization measurements, the error bars are hardly visible in the plot. The interval of covered phase angle is large, between 36$^\circ$ and 116$^\circ$. Very high values of linear polarization have been measured, steadily increasing with the phase angle, although at the highest values it seems that the phase-polarization curve tends to become shallower, as expected for an object approaching $P_{\rm max}$, the value of maximum positive polarization. We note that, as shown by the right Panel of the same Figure, the Calern and Rozhen measurements, obtained independently using two different instruments, agree with each other in the R colour, and at the same time 
the V and R measurements also agree closely. The same is true for all the multi-colour measurements shown in Fig.~\ref{fig:BVRICalern}. It is important to note that there is a regular pattern in the measurements obtained in different colours. In particular, although this is not very clear in Fig.~\ref{fig:BVRICalern} because of the scale of the polarization axis, the measured linear polarization exhibits a slight decrease for decreasing wavelength, from I to B. This follows a pattern which is known to characterize the positive polarization branch of low-albedo asteroids \citep{Belskayaetal05, Bagnuloetal14}.

In order to obtain estimates of some of the relevant polarimetric parameters, we computed a best-fit curve of the available 
phase-polarization data in V, using the trigonometric representation of the phase-polarization relationship of asteroids 
originally proposed by \citet{LummeMuinonen93}, and lately adopted by other authors, including, for instance, 
\citet{Penttilaetal05}:\\
\begin{equation}
P_{\rm r}\left(\alpha\right) = A \sin^B\left(\alpha\right) \cdot \cos^C\left(\frac{\alpha}{2}\right) \cdot \sin\left(\alpha-\alpha_{inv}\right)
\end{equation}
\noindent where $P_{\rm r}\left(\alpha\right)$ is the fraction of linear polarization (with a sign which is 
negative in the negative polarization branch, see, for instance, \citet{MNRAS1}),
$\alpha$ is the phase angle, $\alpha_{inv}$ is the value of the inversion angle of polarization, and $A$, $B$ and $C$
are coefficients to be computed by means of least-squares fits.
This mathematical representation of the phase-polarization curves of atmosphereless Solar system bodies is
eminently empirical (Muinonen, private communication). It was found to provide very reasonable fits of the
phase-polarization curves of objects observed over wide intervals of phase angle, much wider than those
possible for main belt asteroids (which cannot be observed at phase angles much larger than about 30$^\circ$).

The best-fit representation of our V data is shown in the left Panel of Fig.~\ref{fig:phasepolcurveV+R}.
It can be seen that the computed curve fits very well the available data, including, in addition to
our Calern data, also one single measurement available in the literature, obtained at a phase angle of 23$^\circ$
by \citet{Fornasier}. The computed chi-square value of the fit is 0.19. Some noise can also be
due to a mild periodic variation of the linear polarization, that we have found to be synchronous with the rotation period and therefore most likely due to some surface heterogeneity. This is the subject of a separate paper (Borisov et al., in preparation).  
The value of the resulting inversion angle and the 
rms residuals for the whole data set are given in Table \ref{tab:rms}.
By looking at Fig.~\ref{fig:phasepolcurveV+R}, a few considerations can be immediately done: first, although the best-fit model for the V data using the trigonometric representation is quite good, the range of phase angles covered by available measurements is not sufficient to allow us to obtain a very robust determination of the value $P_{\rm max}$ of the maximum positive polarization. On the other hand, if the adopted representation of the phase-polarization curve is valid also slightly beyond the maximum value of phase angle covered by our data, then we find that the value of $P_{\rm max}$  should be reached at a phase angle of about 130$^\circ$, and is about 45\%, a very high value, much higher than the $P_{\rm max}$  values determined so far for any of Solar system body for which polarimetric measurements were made.

The second main consideration concerns the derived value of the inversion angle $\alpha_{inv}$. According to our computations,
this value is about 19$^\circ$ for the whole set of Phaethon V measurements. Such a value is hardly consistent with
the F taxonomic class. On the other hand, the main uncertainty affecting this estimate is
due to the fact that our data do not cover the negative polarization branch, and is therefore based on an
extrapolation to smaller values of phase angle of measurements covering the positive polarization branch, only.

\begin{figure*}
\centerline{
\includegraphics[width=75mm]{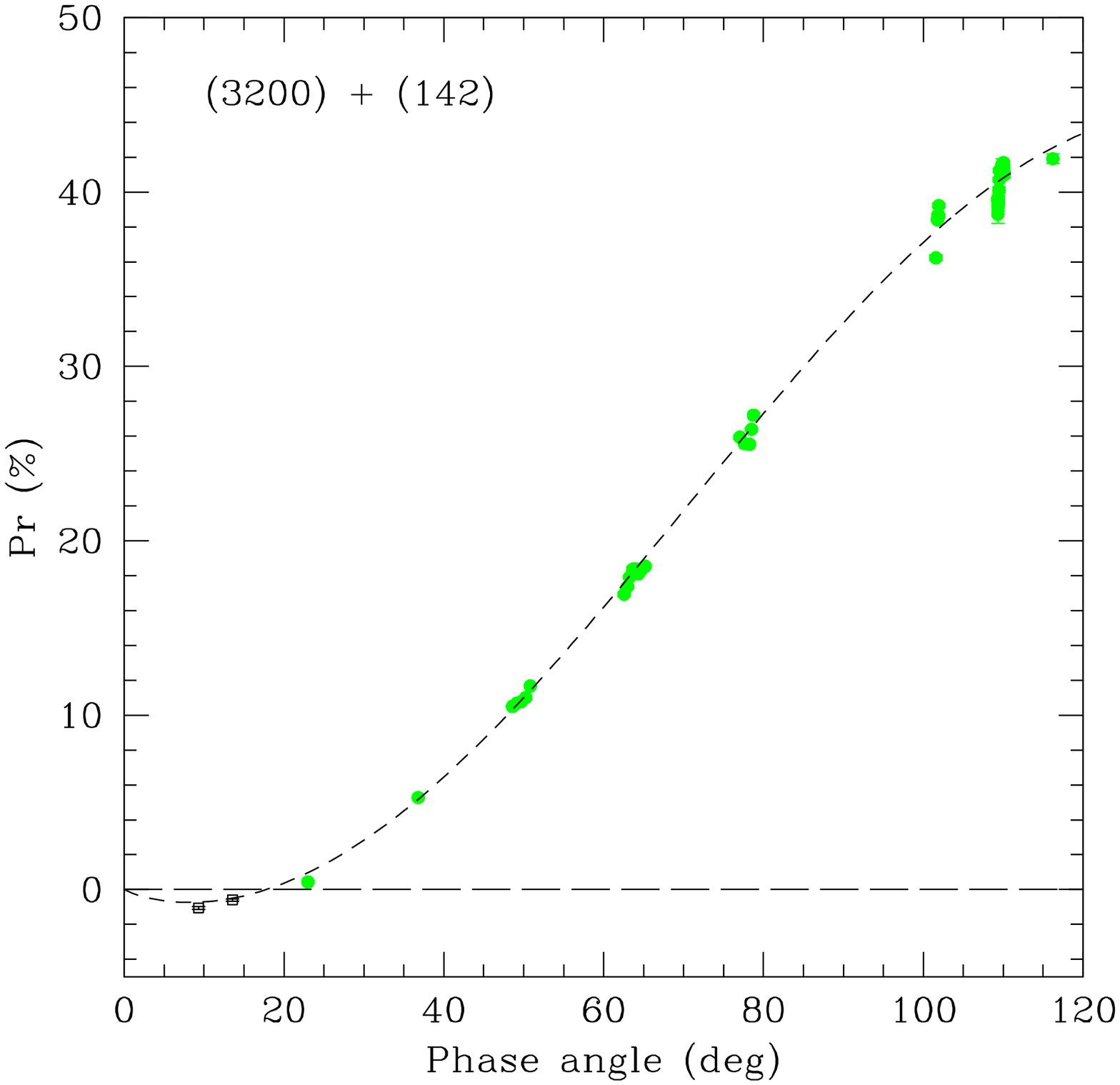}
\includegraphics[width=75mm]{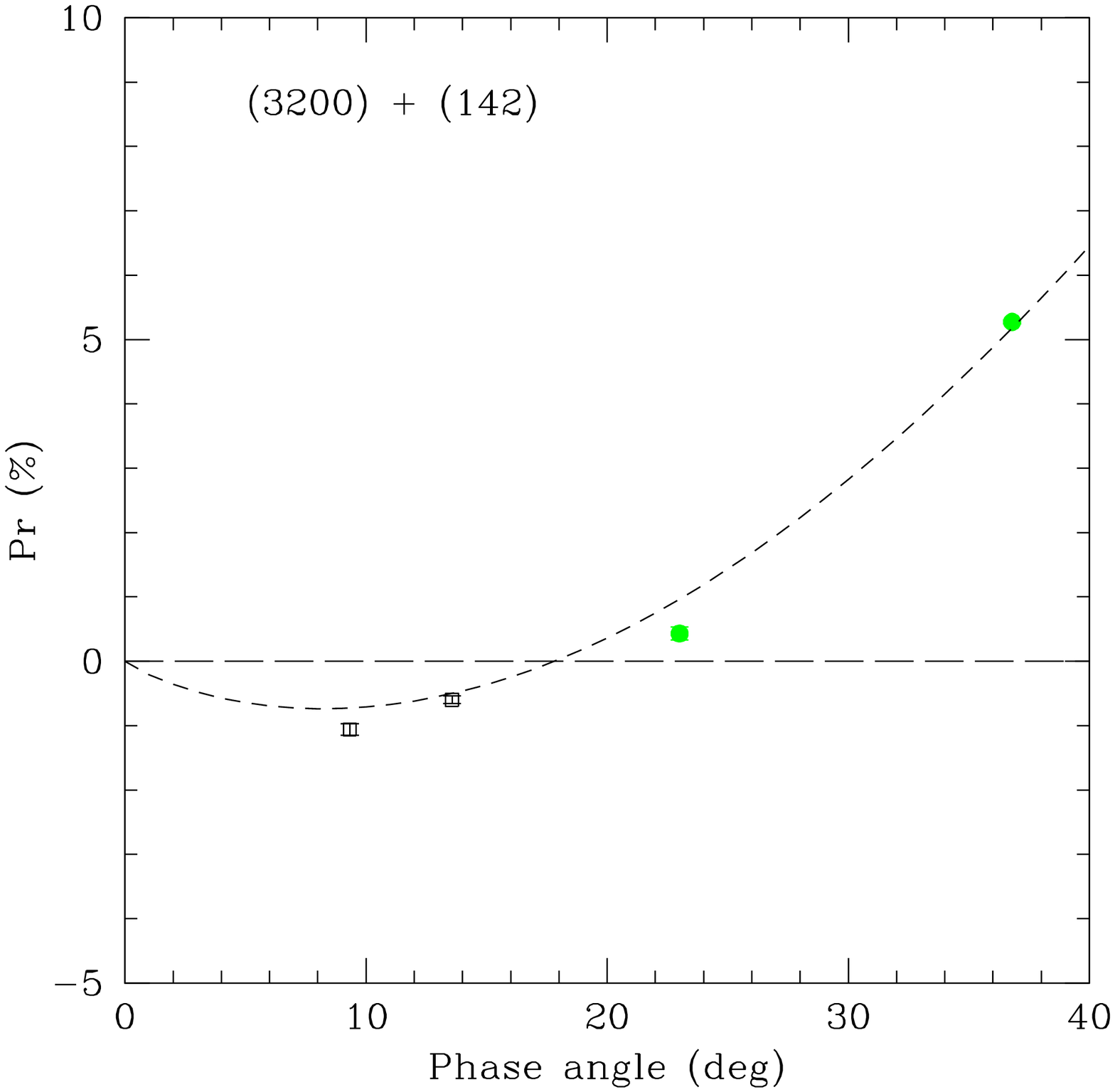} }
\caption{Best-fit phase-polarization curve of the data set obtained by adding to available Phaethon polarization measurements in V light 
just the two measurements available for the F-class asteroid (142)~Polana, both covering the negative polarization branch. The left Panel
shows the full data and the corresponding best-fit. 
For sake of clarity, the right Panel covers only the region of phase angle $<$ 40$^\circ$, where only two measurements 
are available for (3200)~Phaethon (displayed in green).} 
\label{fig:Phaethon_Polana}.
\end{figure*}
\begin{figure*}
\centerline{
\includegraphics[width=75mm]{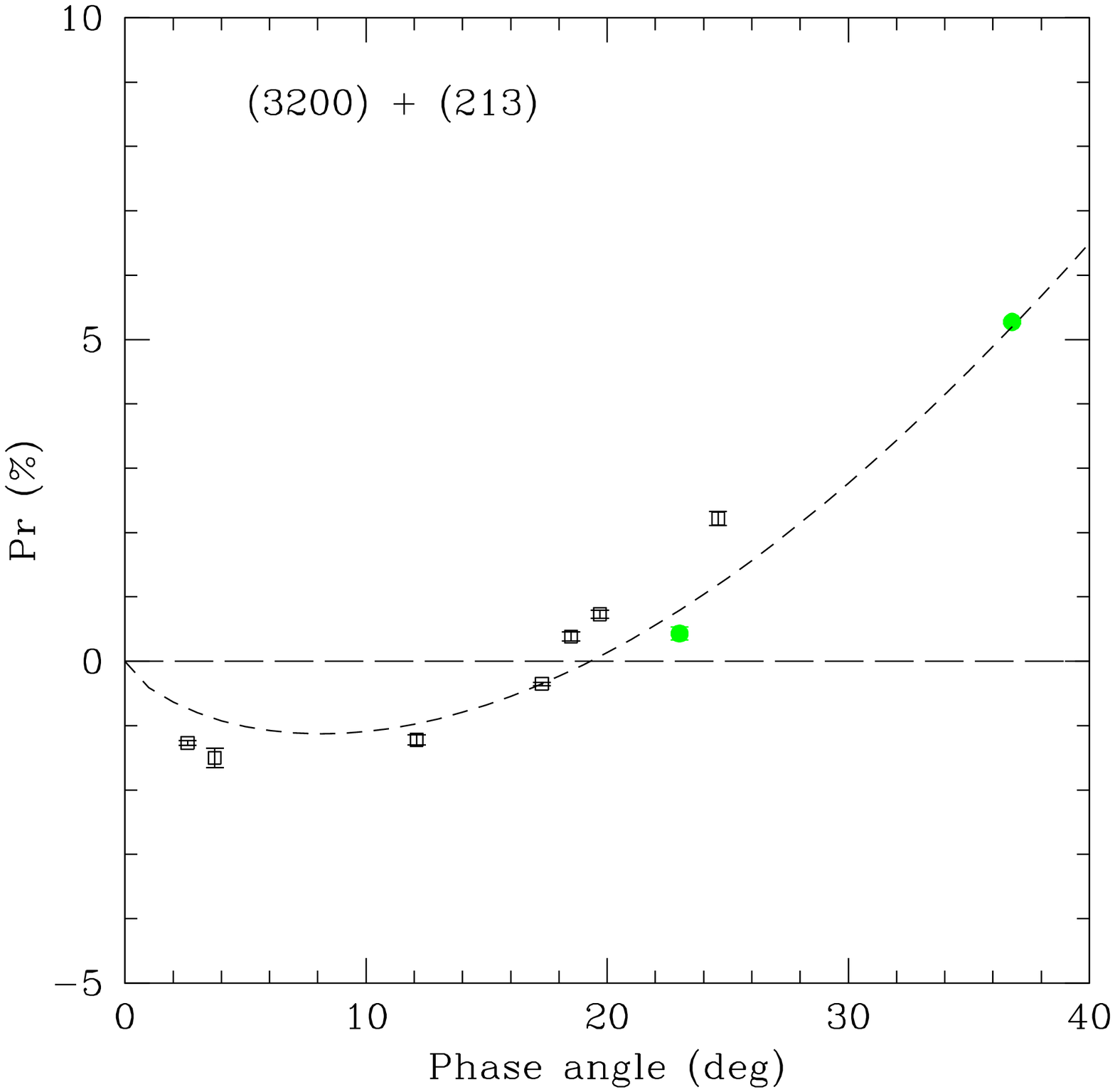}
\includegraphics[width=75mm]{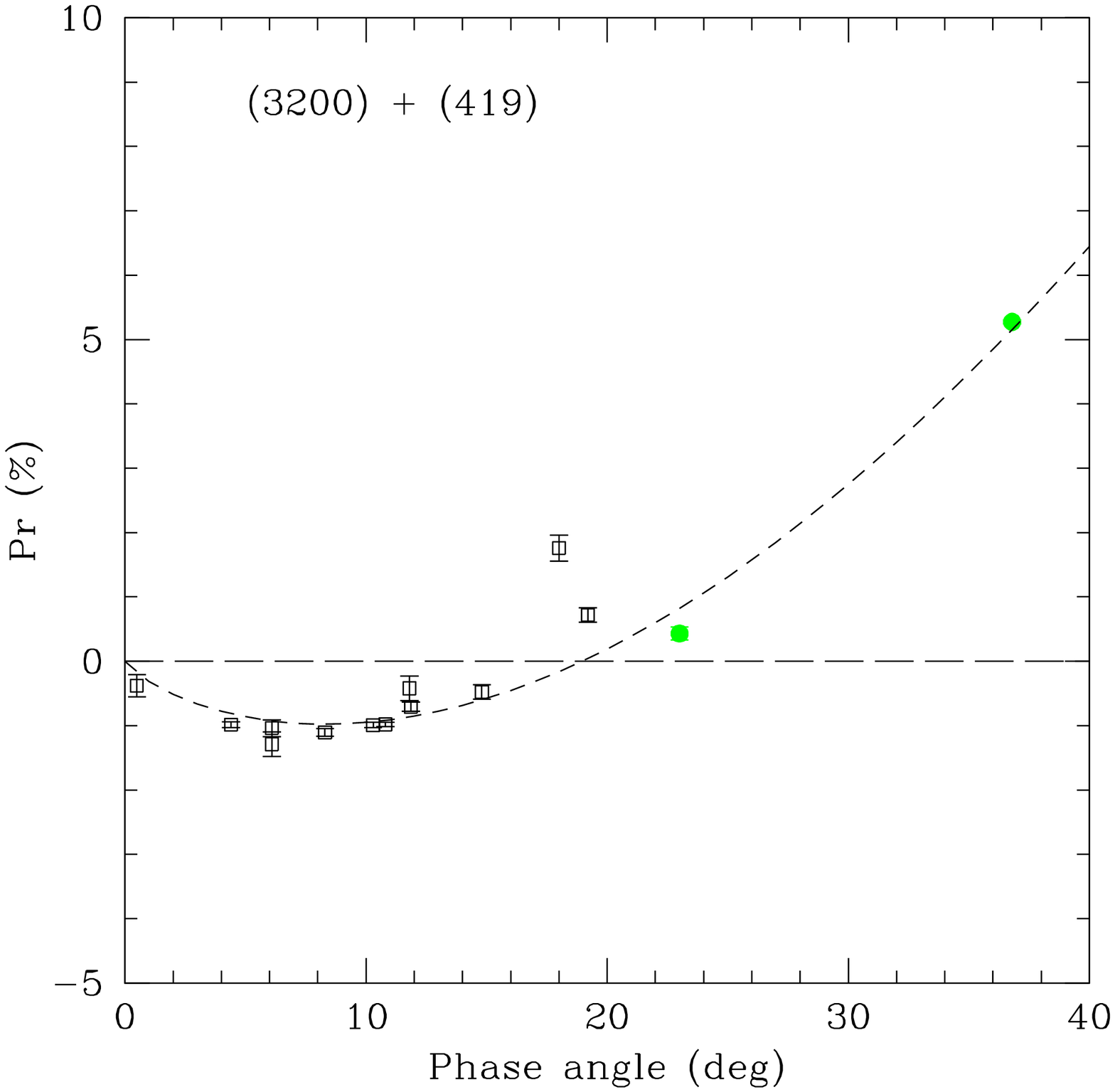} }
\centerline{
\includegraphics[width=75mm]{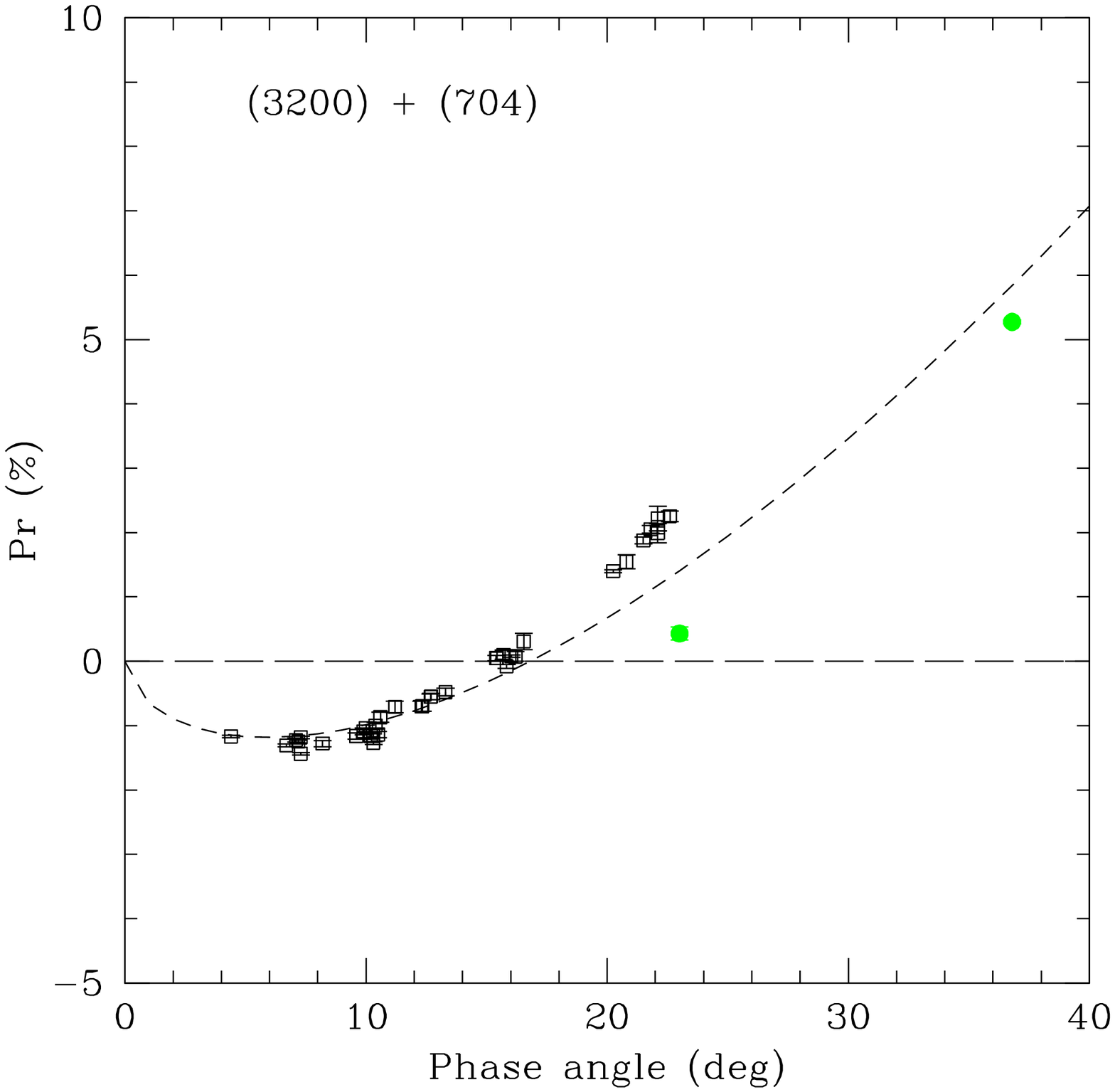}
\includegraphics[width=75mm]{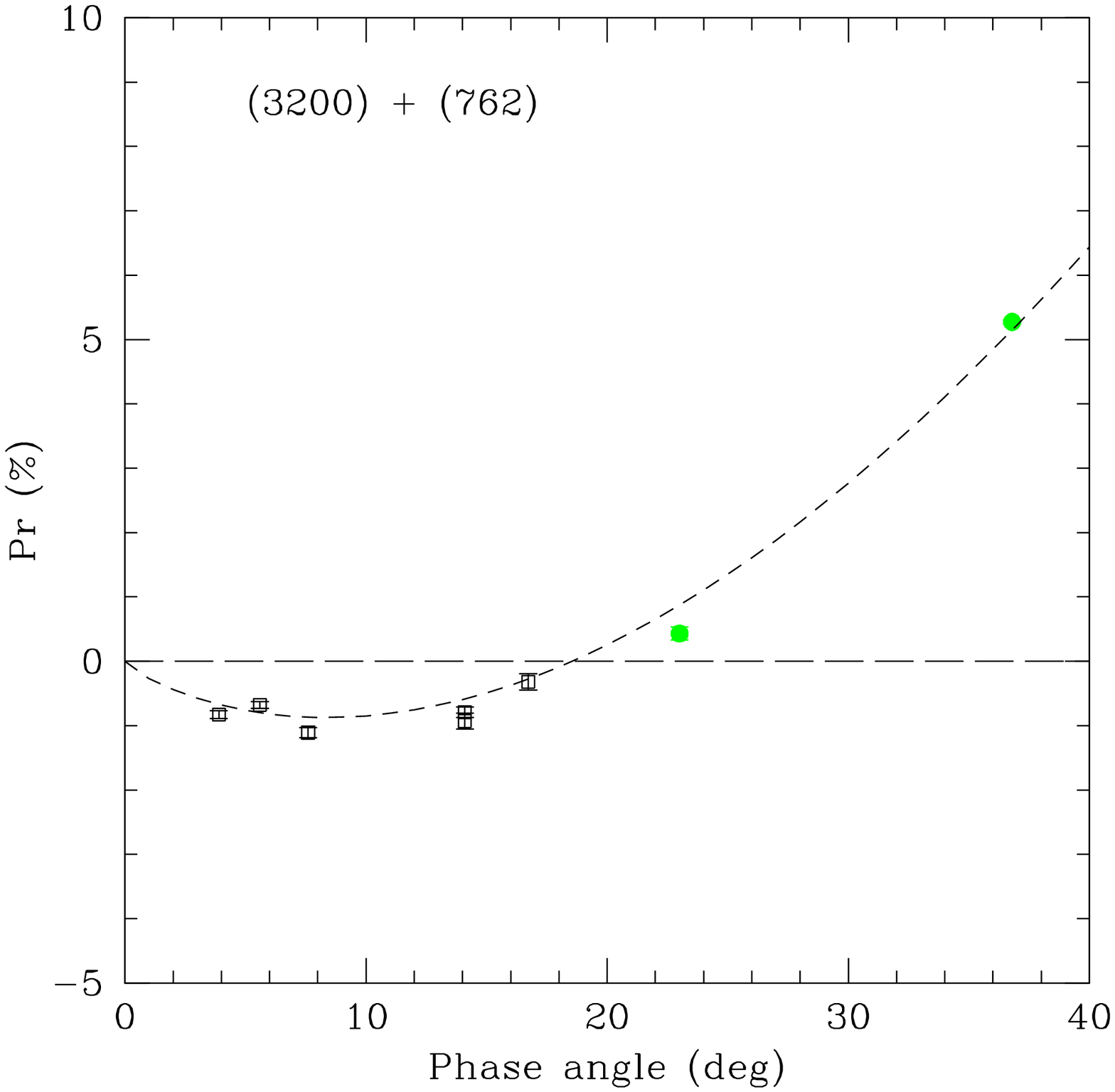} }
\caption{Best-fit phase-polarization curves of the data sets obtained by merging together the (3200)~Phaethon V data shown in 
the left Panel of Fig.~\ref{fig:phasepolcurveV+R} and
available V data for (213)~Lilaea (top left), (419)~Aurelia (top right), (704)~Interamnia (bottom left) and (762)~Pulkova (bottom right).
All these asteroids belong to the old F-class. For sake of clarity, the displayed plots cover only the region 
of phase angle $<$ 40$^\circ$, as in the right Panel of Fig.~\ref{fig:Phaethon_Polana}. Only two measurements 
of (3200)~Phaethon are available in this phase interval, and they are displayed using green symbols. 
The resulting best-fit curves have been computed in each case using all available data, including also the large number of
Phaethon data not shown in the plots.} 
\label{fig:Phaethon_213_419_704_762}.
\end{figure*}

\begin{table}
	\centering
	\caption{Results of the best-fit of available phase-polarimetric data obtained in V light
	for (3200)~Phaethon alone, or for a combination of data for Phaethon and other known
	B-class asteroids, some of them having been classified in the past as members of the old F-class.}
	\label{tab:rms}
	\begin{tabular}{lrcc} 
		\hline
		Data set & \multicolumn{1}{c}{Number of}    & Inversion angle & rms\\
		         & \multicolumn{1}{c}{measurements} & (deg)           & (\%)\\
		\hline
		(3200) alone    &  45 & 18.8 $\pm$ 1.6 & 0.715 \\
    (3200) + (2)    &  73 & 19.2 $\pm$ 0.3 & 0.606 \\
    (3200) + (24)   &  57 & 19.9 $\pm$ 0.6 & 0.658 \\
    (3200) + (47)   &  58 & 19.2 $\pm$ 0.5 & 0.652 \\
    (3200) + (59)   &  55 & 19.4 $\pm$ 0.4 & 0.655 \\
    (3200) + (142)  &  47 & 17.8 $\pm$ 1.2 & 0.699 \\
    (3200) + (213)  &  52 & 19.3 $\pm$ 0.5 & 0.715 \\
    (3200) + (419)  &  57 & 19.0 $\pm$ 0.5 & 0.700 \\
    (3200) + (704)  &  83 & 16.8 $\pm$ 0.5 & 0.769 \\		
    (3200) + (762)  &  51 & 18.5 $\pm$ 0.8 & 0.676 \\
    (3200) + (142), (213),  & 110 & 17.0 $\pm$ 0.4 & 0.690 \\
		(419), (704), (762) &   &       &         \\
		(213) +  (704)  &  45 & 16.2 $\pm$ 0.1 & 0.241 \\
		\hline
	\end{tabular}
\end{table}

Since we seek evidence in favour of, or against, the possibility that Phaethon
belongs to the F taxonomic class, based on its polarimetric properties, and in particular 
on the value of its polarization inversion angle, we computed trigonometric fits of 
sets of measurements obtained by merging together the Phaethon polarization data with those
of a number of asteroids belonging to the modern B-class defined by \citet{BusBinzel},
including also some of the most important members of the old F-class, characterized 
by inversion angles of polarization generally between 16$^\circ$ and 17$^\circ$, significantly smaller than those of the B-class. In particular, we considered asteroids (2)~Pallas,
(24)~Themis, (47)~Aglaja, (59)~Elpis, as representatives of non-F objects belonging to the modern B-class, as well as some members of the old F-class, namely (142)~Polana, (213)~Lilaea, (419)~Aurelia, (762)~Pulkova, and (704)~Interamnia. All the above asteroids are classified as B-type by \citet{BusBinzel}, with the two exceptions of (419) and (762), for which there is no \citet{BusBinzel} classification. We note also that, in the most recent taxonomic classification by \citet{Demeo}, both (2)~Pallas and (3200)~Phaethon are classified as members of the modern B-class, whereas (24)~Themis is classified as C-type. It is interesting to note that (762)~Pulkova is a known binary asteroid, formed by a primary component being about 140 km in size, and a much smaller satellite. Finally, we note that the above list of objects includes asteroids
of quite different sizes. The biggest ones being (2)~Pallas, (24)~Themis and (704)~Interamnia,
with diameters above 500, 200 and 300 km, respectively, while some other
objects, like (142)~Polana, are smaller than 60 km \citep{Masiero11}.

For all selected asteroids we used all polarimetric measurements available in the
literature\footnote{The data come mostly from the Planetary Data System repository, available at the URL address 
http://pds.jpl.nasa.gov/ (files maintained by D.F. Lupishko and I.N. Belskaya), and from some recently published articles, 
including \citep{Gil_2014, Belskayaetal17,Dev_2017}}, plus a few 
still unpublished data obtained at Calern in the the framework of the 
Calern Asteroid Polarimetric Survey (CAPS). These measurements will be included in a new CAPS data release 
to be published in a paper currently in preparation.

\begin{figure}
\includegraphics[width=75mm]{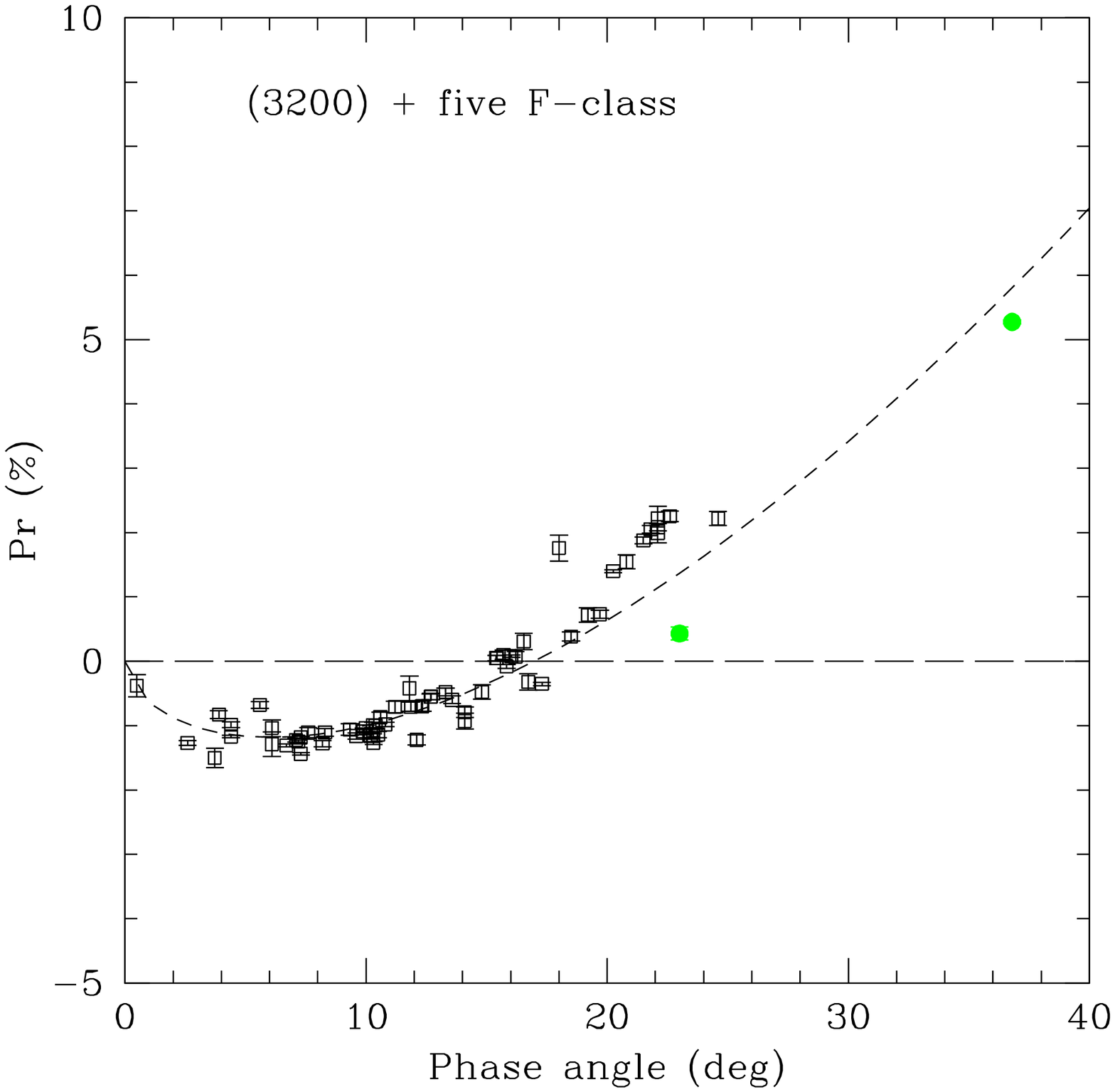} 
\caption{The same as Fig.~\ref{fig:Phaethon_213_419_704_762}, but here the (3200)~Phaethon V data are merged together with all V data available for the F-class asteroids (142)~Polana, (213)~Lilaea, (419)~Aurelia, (704)~Interamnia and (762)~Pulkova, shown separately in Figs.~\ref{fig:Phaethon_Polana} and \ref{fig:Phaethon_213_419_704_762}.} 
\label{fig:Phaethon_Fclass}.
\end{figure}

\begin{figure*}
\centerline{
\includegraphics[width=75mm]{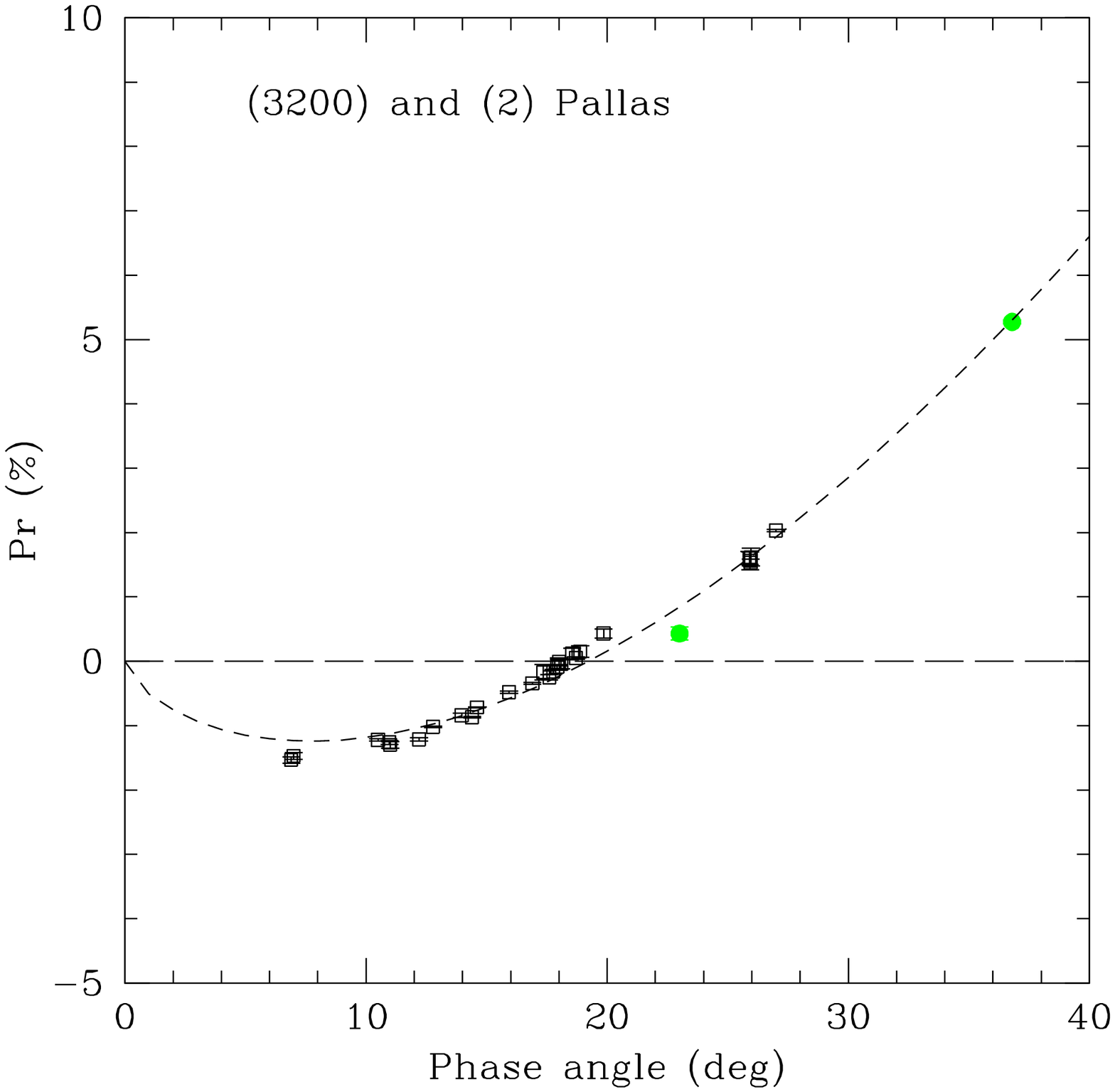}
\includegraphics[width=75mm]{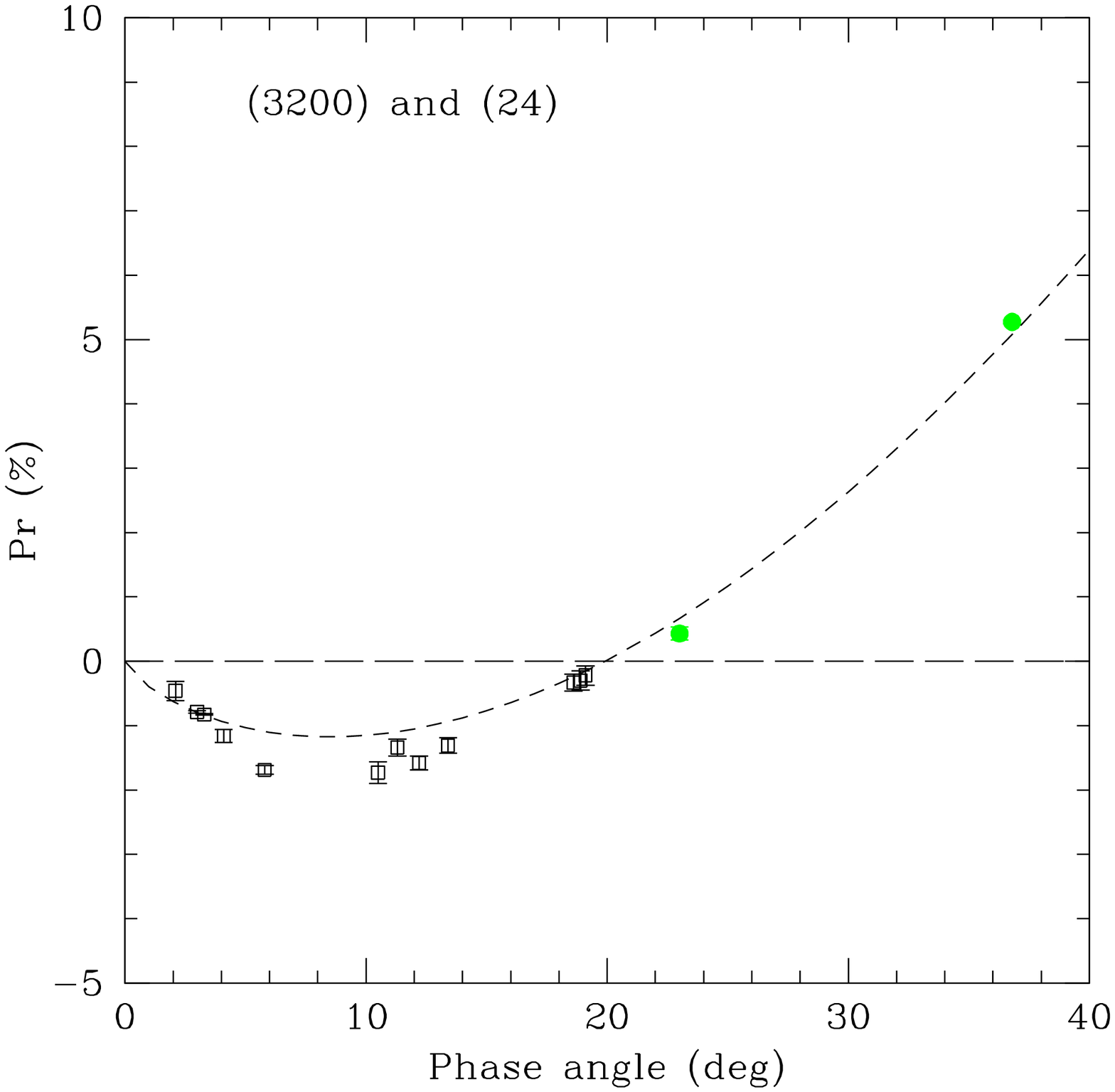} }
\centerline{
\includegraphics[width=75mm]{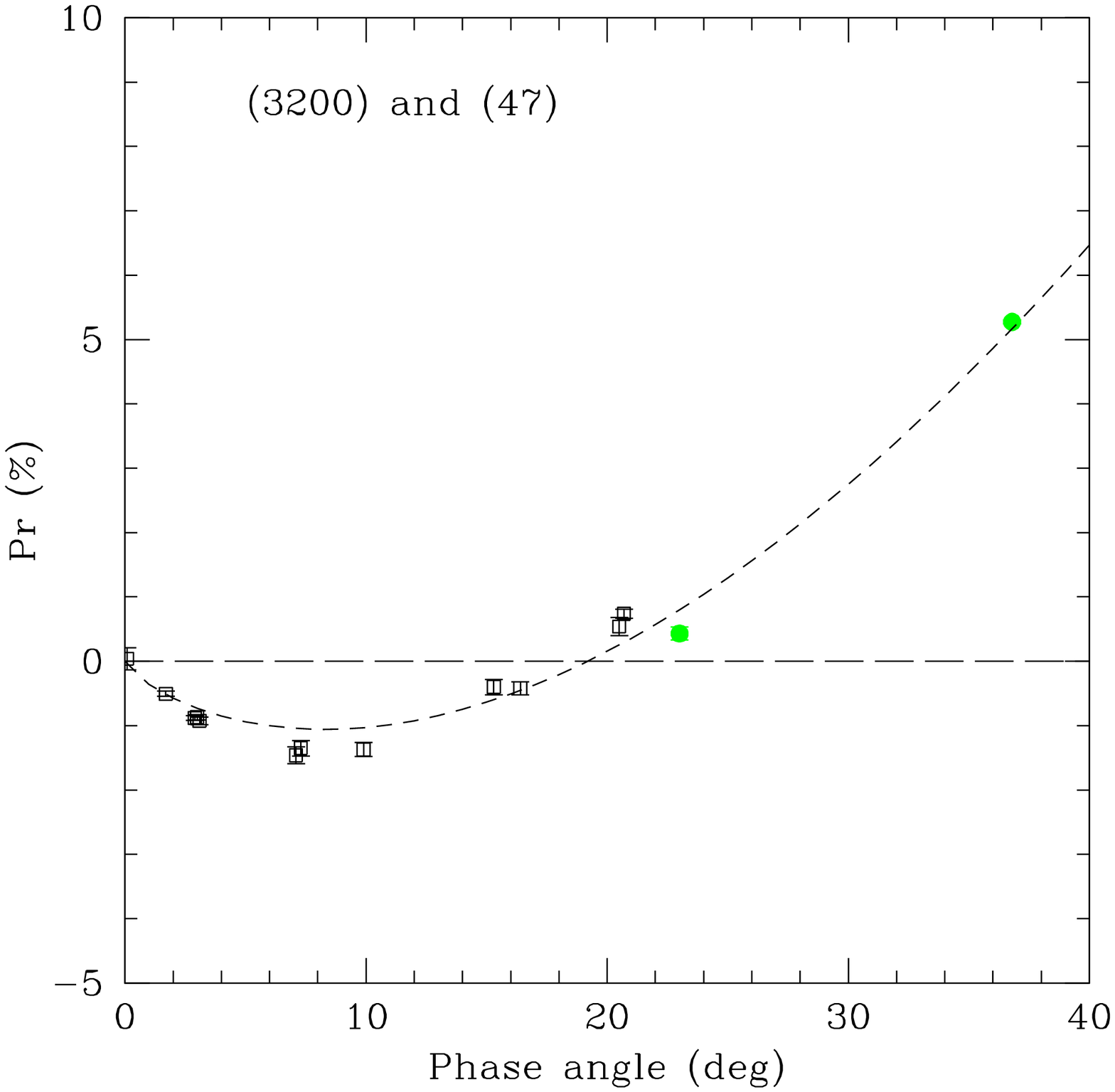}
\includegraphics[width=75mm]{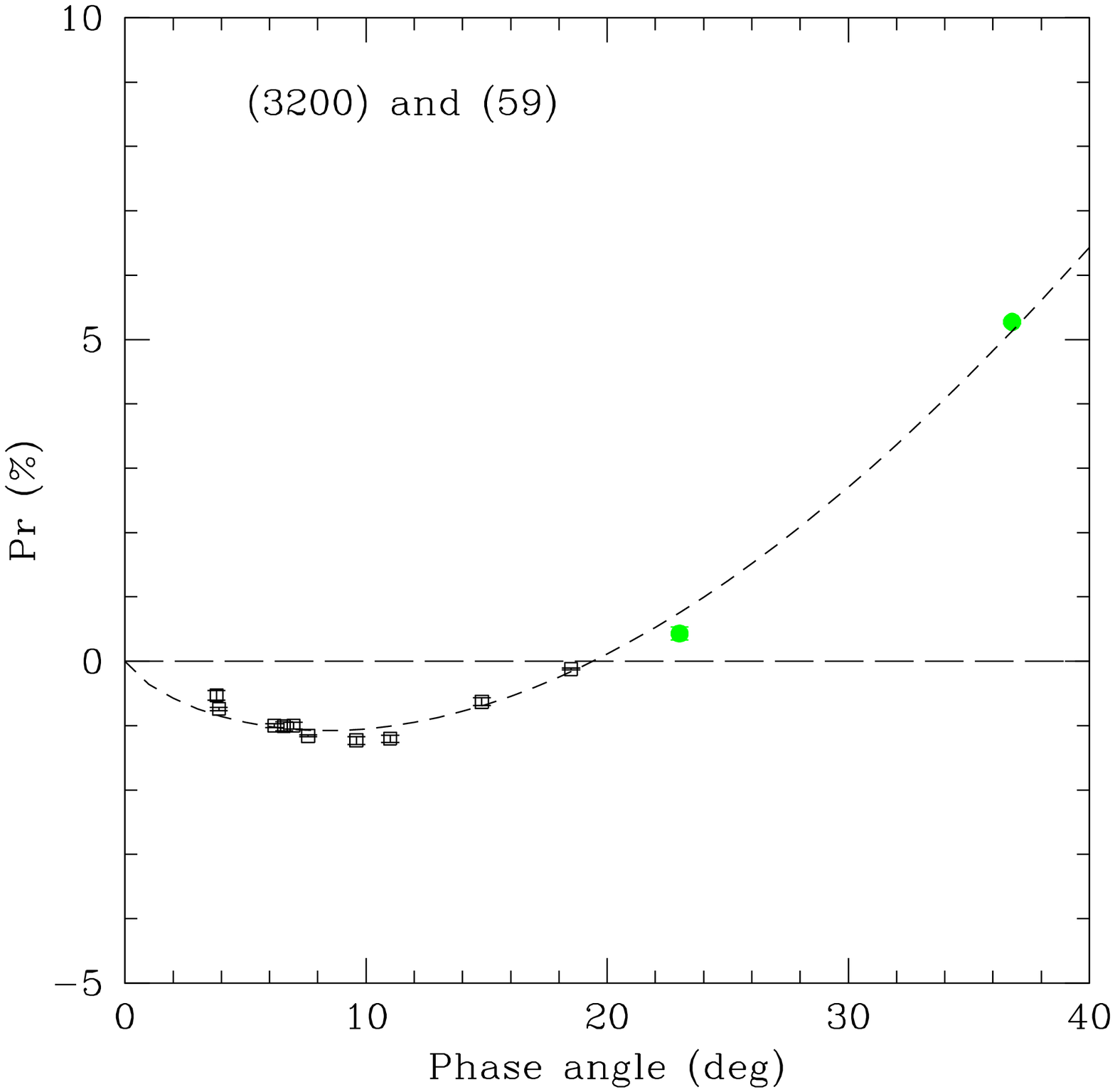} }
\caption{
The same as Fig.~\ref{fig:Phaethon_213_419_704_762}, but here Phaethon polarization data in V light are merged with those for the asteroids (2)~Pallas, (23)~Themis, (47)~Aglaja and (59)~Elpis, all belonging to the modern B-class, but not to the old F class.} 
\label{fig:Phaethon_2_24_47_59}.
\end{figure*}

The results of our exercise are listed in Table~\ref{tab:rms} and graphically shown in Figs.~\ref{fig:Phaethon_Polana}-\ref{fig:Phaethon_2_24_47_59}. Fig.~\ref{fig:Phaethon_Polana} shows what happens when we add to our set of Phaethon measurements in V just a couple of data points covering the negative polarization branch. In particular, we use here the only two measurements obtained in the past at the CASLEO observatory (Argentina) for (142)~Polana, an asteroid belonging to the old F-class. The figure shows that, obviously, the best-fit curve is mostly constrained by the large number of Phaethon measurements, covering a very large interval of phase angles. However, as shown in Table \ref{tab:rms}, adding the two Polana data points produces a shift of the resulting inversion angle of polarization of about one degree towards smaller values, from 18.8$^\circ$ to 17.8$^\circ$, producing at the same time a negligible improvement of the rms. An inversion angle of 17.8$^\circ$, however, does not constitute strong evidence that Phaethon is a member of the F-class, because, according to \citet{Belskayaetal17}, one should expect slightly smaller values for the inversion angle of this class. 

It is therefore interesting to compute the best-fit curve using the same trigonometric representation, in cases when one adds to the Phaethon data larger numbers of measurements of F-class asteroids, in order to see the change in the fitted inversion angle of polarization and in terms of rms residuals. The results of this exercise are shown, again, in Table \ref{tab:rms}, and graphically in Fig.~\ref{fig:Phaethon_213_419_704_762} for each of the F-class asteroids (213), (419), (704) and (762), and in Fig.~\ref{fig:Phaethon_Fclass}, in which the data of all the above-mentioned F-class asteroids are merged together. 

By looking at these plots and at the corresponding values obtained in the different cases for the inversion angle and the rms residuals, it seems that the hypothesis that Phaethon can be a member of the F-class is not strongly supported by the data. The resulting curves show that the measurements of F-class asteroids are not particularly well fitted. This is evident for objects like (213), (419) and (704), for which fairly large numbers of measurements are available. The obtained rms residuals confirm this conclusion. In terms of inversion angle, the obtained values are also larger than expected if Phaethon was an F-class asteroid, characterized by a phase-polarization curve fully compatible with those of other objects of the same class. Only by adding to the Phaethon data all the available data of the five F-class objects mentioned above, the resulting best-fit curve reaches an inversion angle of 17$^\circ$, but the rms do not improve and a visual inspection of the results, shown in Fig.~\ref{fig:Phaethon_Fclass}, does not provide convincing evidence that the obtained best-fit is particularly good, although we cannot firmly state that the data of Phaethon are clearly incompatible with those of the F-class. It is clear that, in order to draw definitive conclusions, we need to observe Phaethon when it is visible at phase angles corresponding to the negative polarization branch.

On the other hand, if one repeats the same exercise, but merging this time our Phaethon data with available data for modern B-class but non-F-class, asteroids, the situation seems to change. In particular, as shown in Table \ref{tab:rms} and in Fig.~\ref{fig:Phaethon_2_24_47_59}, that available polarimetric data for (24) Themis and (47) Aglaja are clearly not compatible with the phase-polarization curve of Phaethon. The situation is better in the case of (59) Elpis. The best result is obtained, however, in the case of the data of (2) Pallas. By merging Phaethon and Pallas data together, the rms of the resulting best-fit curve improves significantly, and this is also clearly visible by looking at the top-left Panel of Fig.~\ref{fig:Phaethon_2_24_47_59}. The corresponding inversion angle of polarization turns out to be 19.2$^\circ \pm$ 0.3$^\circ$, a value fully compatible with the one found for the data of Phaethon alone, but with a smaller associated uncertainty.

Our observations allow to estimate polarimetric albedo based on the empirical relationships ``polarimetric slope-albedo'' and ``Pmax-albedo'' (e.g.,  \citet{GeakeDollfus86}). In both cases the polarimetric albedo of Phaethon is estimated to be very low (about 5\%). This value is quite different from the determined radiometric albedo of Phaethon \citet{Hanusetal16}. What is the possible reason of such discrepancy?  First of all, a thorough check of Phaethon's albedo determined from the radiometric data is required. If there are any doubts in rather high surface albedo of Phaethon we should assume that the polarimetric method of albedo determination failed in the case of this asteroid. This could happen if the surface texture of Phaethon is very different from other asteroids for which polarimetric measurements are available, According to laboratory measurements $P_{\rm max}$  depends on the grain size and increases for dust-free surfaces (e.g. \citet{GeakeDollfus86}). However, an absence of regolith layer seems to contradict thermophysical modeling.

\section{Conclusions}

Although not yet sufficient to derive conclusive evidence about the value of the inversion angle
of (3200)~Phaethon, which is a decisive factor in assigning it to the F-class, a characterization that could be considered to strongly support the hypothesis of a cometary origin, the measurements presented in this paper seem rather to suggest that Phaethon does not belong to the F taxonomic class. We plan to confirm this conclusion by means of further polarimetric measurements that we hope to obtain next December during the next apparition of this object, when it will be observable at phase angles within the negative polarization branch.

For the moment, we can state that, by combining our new polarimetric measurements of Phaethon with those available for main belt asteroids which are members of the modern B class, we find that the best agreement is found when we merge Phaethon polarimetric data with those of (2) Pallas. This seems to confirm the results of previous analyses, in particular those presented by \citet{DeLeonetal10}, reported in Section \ref{whatisit}, which suggested that Phaethon could be a fugitive member of the Pallas dynamical family, presumably produced by a big cratering event on this very large asteroid \citep{Milanietal14}

\begin{figure}
\includegraphics[width=75mm]{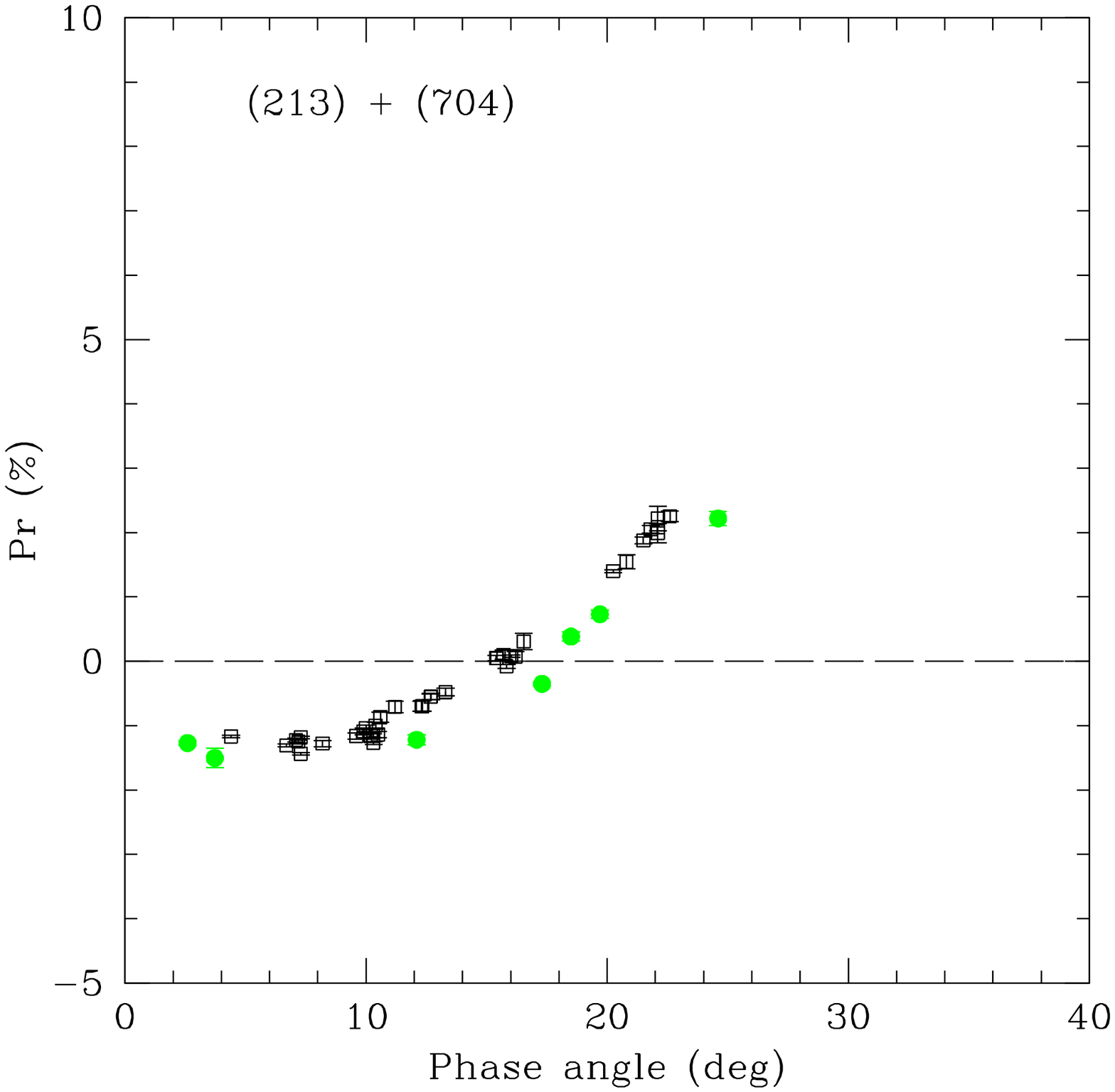}
\caption{Available phase-polarization measurements in V colour for the two F-class asteroids (213)~Lilaea (green symbols) and (704)~Interamnia (black symbols). There are differences between the two objects, but both exhibit an inversion angle of polarization, between about 16$^\circ$ and 18$^\circ$, which is low in comparison to the vast majority of asteroids belonging to other taxonomic classes. } 
\label{fig:213_704}.
\end{figure}

If this conclusion is correct, and confirmed by future data, the problem of the 
origin of Phaethon can be considered solved. On the other hand, other questions
remain open, including the following:
\begin{itemize}
\item What is the real composition of Pallas and its siblings, taking into account
that the geometric albedo of Pallas \citep{MNRAS1}, and also of Phaethon, according to 
\citet{Hanusetal16}, is so high with respect to other asteroids belonging
to the B taxonomic class?
\item Is Pallas rich in volatiles, that can be found also in its siblings, possibly
producing episodes of cometary activity? Is the Pallas family a good candidate to
look for new main-belt comets?
\item What is the origin of Pallas? Was it accreted in its present location, or was it
injected into the asteroid main belt coming from regions at larger heliocentric distance?
\item The reflectance spectrum of Phaethon is very similar to
those of other small members of the Pallas family \citep{DeLeonetal10} which
do not reach the extremely small perihelion distance as Phaethon, and 
are not subject to close encounters with the inner planets. Has this some consequence
about current ideas of the effects of space weathering and resurfacing due to 
tidal effects produced by close encounters with the inner planets?
\end{itemize}
Of course, another possible open question concerns the polarimetric properties of the
old F taxonomic class. In particular, is it possible that this class might be characterized by a larger 
interval of the inversion angle $\alpha_{inv}$ than previously believed? In some cases
this might make it hard to distinguish them from the similar B-class based on 
polarimetric data alone. Fig.~\ref{fig:213_704} shows that for the case of (213)~Lilaea
and (704)~Interamnia, some difference between the phase-polarization curves may exist,
but in any case the inversion angle is small for both objects.
The possibility of a wider heterogeneity in the polarimetric properties of the F-class  would raise some doubts about the predictive power of polarization measurements. This potential problem can be assessed only by means of a survey of spectroscopic measurements extending into the blue region of the reflectance spectrum for objects found to be characterized by a small value of the inversion angle of polarization, in order to associate more closely the polarimetric and spectral reflectance properties of the F-class. Hopefully, this is something that will be achieved by examining the next data releases of Gaia observations of Solar system objects, when spectroscopic data, including also the blue region of the reflectance spectrum, will be included for progressively increasing numbers of asteroids. In the mean time, we will continue our Calern Asteroid Polarimetric Survey program, and will increase the data set of asteroid polarimetric data. 

\section*{Acknowledgements}
\label{sec:Ackn}

The Torino polarimeter was built at the INAF - Torino Astrophysical Observatory 
using dedicated funds provided in the framework of the INAF PRIN 2009 program.
This work is based on data obtained at the C2PU facility (Calern observing station, 
Observatoire de la C\^ote d'Azur, Nice, France) and at the Rozhen National Astronomical Observatory (Bulgaria). 
GB, ZD and YK gratefully acknowledge observing grant support from the Institute of Astronomy and Rozhen National Astronomical Observatory, Bulgarian Academy of Sciences.





\begin{thebibliography}{99}
\bibitem[\protect\citeauthoryear{Bagnulo et al.}{2014}]{Bagnuloetal14}
Bagnulo S., Cellino A., Sterzik M.~F., 2014, MNRAS Letters, 446, L11
%
\bibitem[\protect\citeauthoryear{Bagnulo et al.}{2011}]{Bagnuloetal11}
Bagnulo S. et al., 2011, JQSRT, 112, 2059
%
\bibitem[\protect\citeauthoryear{Bagnulo et al.}{2010}]{Bagnuloetal10}
Bagnulo S. et al., 2010, A\&A, 514, A99
%
\bibitem[\protect\citeauthoryear{Bagnulo et al.}{2009}]{Bagnuloetal09}
Bagnulo S. et al., 2009, PASP, 121, 993
%
\bibitem[\protect\citeauthoryear{Belskaya et al.}{2017}]{Belskayaetal17}
Belskaya I.~N. et al., 2017, Icarus, 284, 30
%
\bibitem[\protect\citeauthoryear{Belskaya et al.}{2005}]{Belskayaetal05}
Belskaya I.~N. et al., 2005, Icarus, 178, 213
%
\bibitem[\protect\citeauthoryear{Bus and Binzel}{2002}]{BusBinzel} 
Bus S.~J. and Binzel R.~P., 2002, Icarus, 158, 146
%
\bibitem[\protect\citeauthoryear{Bowell et al.}{1992}]{Bowelletal92}
Bowell E.~B. et al., 1992, IAU Circular 5585
%
\bibitem[\protect\citeauthoryear{Boehnhardt et al.}{2008}]{Boehnhardt08}
Boehnhardt et al., A\&A, 489, 1337
%
\bibitem[\protect\citeauthoryear{Cellino et al.}{2015a}]{Cellinoetal15}
Cellino A., Gil-Hutton R., Belskaya, I.~N., 2015, In: Polarimetry of Stars 
and Planetary Systems (Kolokolova, L., Levasseur-Regourd, A.-C., Hough, J., Eds.), 
360, Cambridge University Press, Cambridge, U.K
%
\bibitem[\protect\citeauthoryear{Cellino et al.}{2015b}]{MNRAS1}
Cellino A. et al., 2015, MNRAS, 451, 3473
%
\bibitem[\protect\citeauthoryear{Cellino et al.}{2016}]{MNRAS2}
Cellino A. et al., 2016, MNRAS, 455, 2091
%
\bibitem[\protect\citeauthoryear{Chamberline et al.}{1996}]{Chamberlinetal96}
Chamberline A.~B., Mc Fadden L.~A., Schultz R., Schleicher, D.~G., Bus S.~J.,
1996, Icarus, 119, 173
%
\bibitem[\protect\citeauthoryear{Ciarniello et al.}{2015}]{Ciarnielloetal15}
Ciarniello M. et al., 2015, A\&A, 583, A31
%
\bibitem[\protect\citeauthoryear{Cochran and Barker}{1984}]{CochranBarker84}
Cochran A.~L. and Barker E.~S., 1984, Icarus, 59, 296 
%
\bibitem[\protect\citeauthoryear{De Le\'on et al.}{2010}]{DeLeonetal10}
De Le\'on J., Campins H., Tsiganis K., Morbidelli A., Licandro J., 2010,
A\&A, 513, A26
%
\bibitem[\protect\citeauthoryear{DeMeo and Binzel}{2008}]{DeMeoBinzel08}
DeMeo F.~E. and Binzel R.~P., 2008, Icarus, 194, 436
%
\bibitem[\protect\citeauthoryear{DeMeo et al.}{2009}]{Demeo} 
DeMeo F.~E., Binzel R.~P., Slivan S.~M. and Bus S.~J., 2009, Icarus, 202, 160
%
\bibitem[\protect\citeauthoryear{Devog\`{e}le et al}{2017}]{Dev_2017} 
Devog\`{e}le M. et al., 2017, MNRAS, 465, 4335
%
\bibitem[\protect\citeauthoryear{Fern\'andez et al.}{2005}]{Fernandezetal2005}
Fern\'andez J., Jewitt D.~C., Sheppard S.~S., 2005, AJ, 130, 308  
%
\bibitem[\protect\citeauthoryear{Fornasier et al.}{2006}]{Fornasier}
Fornasier S. et al., 2006, A\&A, 455, 371
%
\bibitem[\protect\citeauthoryear{Gustafson}{1989}]{Gustafson89}
Gustafson B~A.~S., 1989, A\&A, 225, 533
%
\bibitem[\protect\citeauthoryear{Gaffey et al.}{1989}]{Gaffeyetal89}
Gaffey M.~J., Bell J.~F., and Cruikshank D.~P.,, 1989, in Asteroids II
(R.~P. Binzel, T. Gehrels, M.~S. Matthews, Eds.), 98, University of Arizona Press, Tucson
%
\bibitem[\protect\citeauthoryear{Geake and Dollfus}{1986}]{GeakeDollfus86}
Geake J. E. and Dollfus A., MNRAS, 218, 75 
%
\bibitem[\protect\citeauthoryear{Gil-Hutton et al.}{2014}]{Gil_2014} 
Gil-Hutton R. et al., A\&A, 569, A122
%
\bibitem[\protect\citeauthoryear{Gradie and Tedesco}{1982}]{GradieTedesco82}
Gradie J. and Tedesco E.~F., 1982, Science, 216, 1405
%
\bibitem[\protect\citeauthoryear{Halliday}{1988}]{Halliday88}
Halliday I., 1988, Icarus, 76, 279
%
\bibitem[\protect\citeauthoryear{Hanu{\v s} et al.}{2016}]{Hanusetal16}
Hanu{\v s} J. et al., 2016, A\&A, 592, A34
%
\bibitem[\protect\citeauthoryear{Hsieh and Jewitt}{2006}]{HsiehJewitt06}
Hsieh H.~H. and Jewitt D., 2006, Science, 312, 561
%
\bibitem[\protect\citeauthoryear{Jewitt and Hsieh}{2006}]{JewittHsieh06}
Jewitt D. and Hsieh H.~H., 2006, AJ, 132, 1624
%
\bibitem[\protect\citeauthoryear{Jockers et al.}{2000}]{Jockersetal2000}
Jockers K. et al., 2000, Kinematika i Fizika Nebesnykh Tel, Suppl., 3, 13
%
\bibitem[\protect\citeauthoryear{Kolokolova and Jockers}{1997}]{KolokoloJockers}
Kolokolova L. and Jockers K., 1997, Planet. Space Sci., 45, 1543
%
\bibitem[\protect\citeauthoryear{Lumme and Muinonen}{1993}]{LummeMuinonen93}
Lumme K. and Muinonen K., 1993, 160 Symposium IAU, Asteroids, Comets, Meteors, 194
%
\bibitem[\protect\citeauthoryear{Masiero at al.}{2011}]{Masiero11}
Masiero J.~R. et al., 2011, ApJ, 749, A104
%
\bibitem[\protect\citeauthoryear{Milani at al.}{2014}]{Milanietal14}
Milani A., et al., 2014, Icarus, 239, 46
%
\bibitem[\protect\citeauthoryear{Oliva}{1997}]{Oliva} 
Oliva E., 1997, A\&ASS, 123, 589
%
\bibitem[\protect\citeauthoryear{Penttila at al.}{2005}]{Penttilaetal05}
Penttila A., Lumme K., Hadamcik E., Levasseur-Regourd A.~C., 2005, A\&A, 432, 1081
%
\bibitem[\protect\citeauthoryear{Pernechele at al}{2012}]{SPIE2012}
Pernechele C., Abe L., Bendjoya Ph., Cellino A., Massone 
G., Tanga P., 2012, Proceedings of the SPIE, 8446, 84462H
%
\bibitem[\protect\citeauthoryear{Shevchenko and Tedesco}{2006}]{ShevTed}
Shevchenko V.~G. and Tedesco E.~F., 2006, Icarus, 184, 211
%
\bibitem[\protect\citeauthoryear{Tedesco et al.}{2002}]{Tedescoetal2002}
Tedesco E.~F., Noah P.~V., Noah M., and Price S.~D., 2002, AJ, 123, 1056
%
\bibitem[\protect\citeauthoryear{Tholen}{1985}]{Tholen85}
Tholen D.~J., 1985, IAU Circular 4034
%
\bibitem[\protect\citeauthoryear{Tholen}{1984}]{Tholen84}
Tholen D.~J., 1984, PhD thesis, University of Arizona
%
\end{thebibliography}




\appendix

\section{Observation data}

\begin{table*}
\caption{Observations of (3200)~Phaethon obtained at the Calern and Rozhen observatories. The dates of observation refer to the beginning of the night, and the epochs of observation are given as Modified Julian Date (JD - 2400000.5). The Filter column gives the Johnson-Cousins filter used in each measurement. Phase is the phase angle of Phaethon at the middle of the observation, $P_{\rm r}$ is the measured polarization, and Observatory corresponds to the observatory were the data were obtained.}
\centering                          
\begin{tabular}{r c c c c c}          
\hline\hline                 
 Date           & MJD           &  Filter & Phase                       & $P_{\rm r}$ & Observatory \\ 
                &               &         & \multicolumn{1}{c}{[Deg]}   & [\%]        &           \\  
\hline                        
14/12/2017 & 58101.755 & V & 36.79 & 5.27 $\pm$ 0.03 & Calern \\
14/12/2017 & 58101.759 & R & 36.85 & 5.45 $\pm$ 0.03 & Calern\\
14/12/2017 & 58101.833 & B & 37.53 & 5.08 $\pm$ 0.03 & Calern\\
14/12/2017 & 58101.766 & I & 36.92 & 5.71 $\pm$ 0.02 & Calern \\
15/12/2017 & 58102.696 & R & 47.48 & 9.89 $\pm$ 0.06 & Rozhen\\
15/12/2017 & 58102.703 & R & 47.55 & 9.86 $\pm$ 0.06 & Rozhen\\
15/12/2017 & 58102.707 & R & 47.63 & 9.89 $\pm$ 0.07 & Rozhen\\
15/12/2017 & 58102.712 & R & 47.69 & 10.01 $\pm$ 0.07 & Rozhen\\
15/12/2017 & 58102.717 & R & 47.76 & 10.04 $\pm$ 0.07 & Rozhen\\
15/12/2017 & 58102.723 & R & 47.83 & 10.03 $\pm$ 0.07 & Rozhen\\
15/12/2017 & 58102.729 & R & 47.91 & 10.07 $\pm$ 0.07 & Rozhen\\
15/12/2017 & 58102.734 & R & 47.98 & 10.07 $\pm$ 0.07 & Rozhen\\
15/12/2017 & 58102.740 & R & 47.98 & 10.09 $\pm$ 0.06 & Rozhen\\
15/12/2017 & 58102.745 & R & 48.12 & 10.10 $\pm$ 0.06 & Rozhen\\
15/12/2017 & 58102.752 & R & 48.20 & 10.18 $\pm$ 0.06 & Rozhen\\
15/12/2017 & 58102.757 & R & 48.28 & 10.32 $\pm$ 0.07 & Rozhen\\
15/12/2017 & 58102.763 & R & 48.36 & 10.24 $\pm$ 0.07 & Rozhen\\
15/12/2017 & 58102.769 & R & 48.43 & 10.40 $\pm$ 0.08 & Rozhen\\
15/12/2017 & 58102.775 & R & 48.52 & 10.54 $\pm$ 0.08 & Rozhen\\
15/12/2017 & 58102.781 & R & 48.59 & 10.46 $\pm$ 0.09 & Rozhen\\
15/12/2017 & 58102.782 & V & 48.59 & 10.50 $\pm$ 0.04 & Calern\\
15/12/2017 & 58102.787 & R & 48.67 & 10.62 $\pm$ 0.10 & Rozhen\\
15/12/2017 & 58102.822 & V & 49.13 & 10.68 $\pm$ 0.03 & Calern\\
15/12/2017 & 58102.792 & R & 48.74 & 10.74 $\pm$ 0.07 & Rozhen\\
15/12/2017 & 58102.798 & R & 48.82 & 10.86 $\pm$ 0.06 & Rozhen\\
15/12/2017 & 58102.804 & R & 48.89 & 10.81 $\pm$ 0.06 & Rozhen\\
15/12/2017 & 58102.810 & R & 47.98 & 10.83 $\pm$ 0.06 & Rozhen\\
15/12/2017 & 58102.816 & R & 49.05 & 10.85 $\pm$ 0.06 & Rozhen\\
15/12/2017 & 58102.822 & R & 49.12 & 10.86 $\pm$ 0.06 & Rozhen\\
15/12/2017 & 58102.828 & R & 49.22 & 10.80 $\pm$ 0.06 & Rozhen\\
15/12/2017 & 58102.834 & R & 49.30 & 10.85 $\pm$ 0.06 & Rozhen\\
15/12/2017 & 58102.840 & R & 49.36 & 10.75 $\pm$ 0.06 & Rozhen\\
15/12/2017 & 58102.847 & R & 49.39 & 10.65 $\pm$ 0.06 & Rozhen\\
15/12/2017 & 58102.853 & R & 49.54 & 10.87 $\pm$ 0.06 & Rozhen\\
15/12/2017 & 58102.866 & V & 49.70 & 10.78 $\pm$ 0.02 & Calern\\
15/12/2017 & 58102.907 & V & 50.26 & 11.01 $\pm$ 0.03 & Calern\\
15/12/2017 & 58102.949 & V & 50.82 & 11.69 $\pm$ 0.03 & Calern\\
15/12/2017 & 58102.786 & R & 48.66 & 10.78 $\pm$ 0.04 & Calern\\
15/12/2017 & 58102.827 & R & 49.19 & 10.90 $\pm$ 0.02 & Calern\\
15/12/2017 & 58102.861 & R & 49.77 & 11.03 $\pm$ 0.04 & Calern\\
15/12/2017 & 58102.912 & R & 50.32 & 11.34 $\pm$ 0.03 & Calern\\
15/12/2017 & 58102.953 & R & 50.87 & 11.90 $\pm$ 0.02 & Calern \\
15/12/2017 & 58102.793 & I & 48.73 & 11.27 $\pm$ 0.02 & Calern\\
15/12/2017 & 58102.836 & I & 49.31 & 11.33 $\pm$ 0.02 & Calern \\
15/12/2017 & 58102.877 & I & 49.85 & 11.55 $\pm$ 0.02 & Calern \\
15/12/2017 & 58102.918 & I & 50.41 & 11.89 $\pm$ 0.05 & Calern\\
15/12/2017 & 58102.960 & I & 50.96 & 12.40 $\pm$ 0.02 & Calern\\
15/12/2017 & 58102.807 & B & 48.91 & 10.19 $\pm$ 0.07 & Calern\\
15/12/2017 & 58102.850 & B & 49.49 & 10.14 $\pm$ 0.07 & Calern\\
15/12/2017 & 58102.892 & B & 50.04 & 10.41 $\pm$ 0.04 & Calern\\
15/12/2017 & 58102.933 & B & 50.59 & 10.95 $\pm$ 0.04 & Calern \\
15/12/2017 & 58102.976 & B & 51.17 & 11.13 $\pm$ 0.04 & Calern\\
16/12/2017 & 58103.783 & V & 62.57 & 16.93 $\pm$ 0.08 & Calern \\
16/12/2017 & 58103.812 & V & 62.99 & 17.37 $\pm$ 0.08 & Calern \\
16/12/2017 & 58103.829 & V & 63.25 & 17.92 $\pm$ 0.08 & Calern \\
16/12/2017 & 58103.857 & V & 63.66 & 18.37 $\pm$ 0.09 & Calern \\
16/12/2017 & 58103.876 & V & 63.92 & 18.39 $\pm$ 0.09 & Calern \\
16/12/2017 & 58103.905 & V & 64.36 & 18.10 $\pm$ 0.09 & Calern \\
16/12/2017 & 58103.922 & V & 64.62 & 18.30 $\pm$ 0.09 & Calern \\
16/12/2017 & 58103.960 & V & 65.18 & 18.54 $\pm$ 0.09 & Calern \\
\hline                                   
\end{tabular}
\label{Tab:data}
\end{table*} 

\begin{table*}
\contcaption{continued}
\centering                          
\addtocounter{table}{-1}
\begin{tabular}{r c c c c c}          
\hline\hline                 
 Date           & MJD           &  Filter & Phase                       & $P_{\rm r}$ & Observatory \\ 
                &               &         & \multicolumn{1}{c}{[Deg]}   & [\%]        &           \\  
\hline                        
16/12/2017 & 58103.789 & R & 62.66 & 17.27 $\pm$ 0.10 & Calern \\
16/12/2017 & 58103.818 & R & 63.08 & 17.87 $\pm$ 0.10 & Calern \\
16/12/2017 & 58103.835 & R & 63.33 & 18.46 $\pm$ 0.10 & Calern \\
16/12/2017 & 58103.862 & R & 63.74 & 18.78 $\pm$ 0.10 & Calern \\
16/12/2017 & 58103.881 & R & 64.03 & 18.56 $\pm$ 0.10 & Calern \\
16/12/2017 & 58103.910 & R & 64.44 & 18.55 $\pm$ 0.10 & Calern \\
16/12/2017 & 58103.931 & R & 64.75 & 18.55 $\pm$ 0.10 & Calern \\
16/12/2017 & 58103.967 & R & 65.29 & 19.12 $\pm$ 0.11 & Calern \\
16/12/2017 & 58103.795 & I & 62.73 & 18.21 $\pm$ 0.07 & Calern \\
16/12/2017 & 58103.824 & I & 63.16 & 18.94 $\pm$ 0.08 & Calern \\
16/12/2017 & 58103.840 & I & 63.41 & 19.37 $\pm$ 0.09 & Calern \\
16/12/2017 & 58103.869 & I & 63.83 & 19.60 $\pm$ 0.06 & Calern \\
16/12/2017 & 58103.888 & I & 64.11 & 19.35 $\pm$ 0.07 & Calern \\
16/12/2017 & 58103.916 & I & 64.52 & 19.65 $\pm$ 0.06 & Calern \\
16/12/2017 & 58103.939 & I & 64.86 & 19.51 $\pm$ 0.05 & Calern \\
16/12/2017 & 58103.804 & B & 62.86 & 16.66 $\pm$ 0.05 & Calern \\
16/12/2017 & 58103.849 & B & 63.51 & 17.85 $\pm$ 0.07 & Calern \\
16/12/2017 & 58103.897 & B & 64.21 & 17.30 $\pm$ 0.06 & Calern \\
16/12/2017 & 58103.950 & B & 65.00 & 17.65 $\pm$ 0.09 & Calern \\
17/12/2017 & 58104.780 & R & 77.12  & 26.65 $\pm$ 0.12& Calern\\
17/12/2017 & 58104.822 & R & 77.72  & 25.04 $\pm$ 0.23 & Calern \\
17/12/2017 & 58104.865 & R & 78.34  & 26.42 $\pm$ 0.10 & Calern\\
17/12/2017 & 58104.882 & R & 78.57  & 27.25 $\pm$ 0.10& Calern\\
17/12/2017 & 58104.900 & R & 78.82  & 27.95 $\pm$ 0.11& Calern\\
17/12/2017 & 58104.802 & B & 77.42  & 24.555  $\pm$ 0.12& Calern\\
17/12/2017 & 58104.844 & B & 78.03 & 24.147 $\pm$ 0.13& Calern\\
17/12/2017 & 58104.787 & I & 77.21  & 27.172 $\pm$ 0.08& Calern\\
17/12/2017 & 58104.829 & I & 77.82 & 26.981  $\pm$ 0.09& Calern\\
17/12/2017 & 58104.871 & I & 78.42 & 27.675 $\pm$ 0.09& Calern\\
17/12/2017 & 58104.889 & I & 78.66 & 28.412 $\pm$ 0.07& Calern \\
17/12/2017 & 58104.775 & V & 77.04 & 25.942 $\pm$ 0.09& Calern\\
17/12/2017 & 58104.816 & V & 77.63 & 25.563 $\pm$ 0.09& Calern\\
17/12/2017 & 58104.861 & V & 78.27 & 25.535 $\pm$ 0.09& Calern\\
17/12/2017 & 58104.878 & V & 78.51 & 26.39  $\pm$0.09& Calern\\
17/12/2017 & 58104.895 & V & 78.76 & 27.201 $\pm$ 0.11& Calern\\
19/12/2017 & 58106.807 & V & 101.57 & 36.229 $\pm$ 0.16& Calern \\
19/12/2017 & 58106.829 & V & 101.78 & 38.402 $\pm$ 0.09& Calern \\
19/12/2017 & 58106.833 & V & 101.81 & 38.539 $\pm$ 0.09& Calern\\
19/12/2017 & 58106.836 & V & 101.84 & 38.684 $\pm$ 0.09& Calern \\
19/12/2017 & 58106.838 & V & 101.87 & 38.587 $\pm$0.10& Calern\\
19/12/2017 & 58106.841 & V & 101.90 & 38.675 $\pm$ 0.09 & Calern\\
19/12/2017 & 58106.846 & V & 101.94 & 39.216 $\pm$ 0.10 & Calern\\
20/12/2017 & 58107.701 & R & 109.25 & 41.17 $\pm$ 0.06 & Rozhen\\
21/12/2017 & 58108.718 & R & 116.25 & 43.50 $\pm$ 0.10 & Rozhen\\
\hline                                   
\end{tabular}
\end{table*} 

%
%
%


\bsp	
\label{lastpage}
\end{document}